\documentclass[twocolumn]{aastex61}
\usepackage{mathtools}
\DeclarePairedDelimiter{\abs}{\lvert}{\rvert}

\received{}
\revised{}
\accepted{}

\shorttitle{HSC $z\sim5$ QLF}
\shortauthors{Niida et al.}

\begin{document}
\title{The faint end of the quasar luminosity function at $z \sim 5$ from the Subaru Hyper Suprime-Cam survey}

\correspondingauthor{Mana Niida}
\email{niida@cosmos.phys.sci.ehime-u.ac.jp}

\author{Mana Niida}
\affil{Graduate School of Science and Engineering, Ehime University, 2-5 Bunkyo-cho, Matsuyama, 
Ehime 790-8577, Japan}

\author{Tohru Nagao}
\affil{Research Center for Space and Cosmic Evolution, Ehime University, 2-5 Bunkyo-cho, Matsuyama, 
Ehime 790-8577, Japan}

\author{Hiroyuki Ikeda}
\affil{National Astronomical Observatory of Japan, Mitaka, Tokyo 181-8588, Japan}
\affil{National Institute of Technology, Wakayama College, 77 Noshima, Nada-cho, Gobo, Wakayama 644-0023, Japan}

\author{Masayuki Akiyama}
\affil{Astronomical Institute, Tohoku University, Aramaki, Aoba-ku, Sendai, 980-8578, Japan}

\author{Yoshiki Matsuoka}
\affil{Research Center for Space and Cosmic Evolution, Ehime University, 2-5 Bunkyo-cho, Matsuyama, 
Ehime 790-8577, Japan}

\author{Wanqiu He}
\affil{Astronomical Institute, Tohoku University, Aramaki, Aoba-ku, Sendai, 980-8578, Japan}

\author{Kenta Matsuoka}
\affil{Dipartimento di Fisica e Astronomia, Universit$\acute{a}$ degli Studi di Firenze, Via G. Sansone 1, 
I-50019 Sesto Fiorentino, Italy}
\affil{Research Center for Space and Cosmic Evolution, Ehime University, 2-5 Bunkyo-cho, Matsuyama, 
Ehime 790-8577, Japan}

\author{Yoshiki Toba}
\affil{Department of Astronomy, Kyoto University, Kitashirakawa-Oiwake-cho, Sakyo-ku, Kyoto 606-8502, Japan}
\affil{Academia Sinica Institute of Astronomy and Astrophysics, 11F of Astronomy-Mathematics Building, 
AS/NTU, No.1, Section 4, Roosevelt Road, Taipei 10617, Taiwan}
\affil{Research Center for Space and Cosmic Evolution, Ehime University, 2-5 Bunkyo-cho, Matsuyama, 
Ehime 790-8577, Japan}

\author{Masafusa Onoue}
\affil{Max Planck Institut f\"{u}r Astronomie, K\"{o}nigstuhl 17, D-69117, Heidelberg, Germany}

\author{Masakazu A. R. Kobayashi}
\affil{Faculty of Natural Sciences, National Institute of Technology, Kure College, 2-2-11, Agaminami, Kure, 
Hiroshima 737-8506, Japan}

\author{Yoshiaki Taniguchi}
\affil{The Open University of Japan, 2-11, Wakaba, Mihama-ku, Chiba 261-8586, Japan}

\author{Hisanori Furusawa}
\affil{National Astronomical Observatory of Japan, Mitaka, Tokyo 181-8588, Japan}

\author{Yuichi Harikane}
\affil{Institute for Cosmic Ray Research, The University of Tokyo, 5-1-5 Kashiwanoha, Kashiwa, 
Chiba 277-8582, Japan}
\affil{Department of Physics, Graduate School of Science, The University of Tokyo, 7-3-1 Hongo, Bunkyo, 
Tokyo 113-0033, Japan}

\author{Masatoshi Imanishi}
\affil{National Astronomical Observatory of Japan, Mitaka, Tokyo 181-8588, Japan}
\affil{Department of Astronomical Science, Graduate University for Advanced Studies (SOKENDAI), Mitaka, 
Tokyo 181-8588, Japan}

\author{Nobunari Kashikawa}
\affil{Department of Astronomy, School of Science, The University of Tokyo, Tokyo 113-0033, Japan}
\affil{National Astronomical Observatory of Japan, Mitaka, Tokyo 181-8588, Japan}
\affil{Department of Astronomical Science, Graduate University for Advanced Studies (SOKENDAI), Mitaka, 
Tokyo 181-8588, Japan}

\author{Toshihiro Kawaguchi}
\affil{Department of Economics, Management and Information Science, Onomichi City University, Onomichi, 
Hiroshima 722-8506, Japan}

\author{Yutaka Komiyama}
\affil{National Astronomical Observatory of Japan, Mitaka, Tokyo 181-8588, Japan}
\affil{Department of Astronomical Science, Graduate University for Advanced Studies (SOKENDAI), Mitaka, 
Tokyo 181-8588, Japan}

\author{Hikari Shirakata}
\affil{Department of Cosmosciences, Graduates School of Science, Hokkaido University, N10 W8, Kitaku, 
Sapporo 060-0810, Japan}

\author{Yuichi Terashima}
\affil{Graduate School of Science and Engineering, Ehime University, 2-5 Bunkyo-cho, Matsuyama, 
Ehime 790-8577, Japan}

\author{Yoshihiro Ueda}
\affil{Department of Astronomy, Kyoto University, Kitashirakawa-Oiwake-cho, Sakyo-ku, Kyoto 606-8502, Japan}

\begin{abstract}
We present the quasar luminosity function at $z \sim 5$ derived from the optical wide-field survey data 
obtained as a part of the Subaru strategic program (SSP) with Hyper Suprime-Cam (HSC). From 
$\sim$81.8 deg$^2$ area in the Wide layer of the HSC-SSP survey, we selected 224 candidates of 
low-luminosity quasars at $z \sim 5$ by adopting the Lyman-break method down to $i = 24.1$ mag. 
Based on our candidates and spectroscopically-confirmed quasars from the Sloan Digital Sky Survey 
(SDSS), we derived the quasar luminosity function at $z \sim 5$ covering a wide luminosity range of 
$-28.76 < M_{\rm 1450} < -22.32$ mag. We found that the quasar luminosity function is fitted by a double 
power-law model with a break magnitude of $M^{*}_{1450} = -25.05^{+0.10}_{-0.24}$ mag. 
The inferred number density of low-luminosity quasars is lower, and the derived faint-end slope, 
$-1.22^{+0.03}_{-0.10}$, is flatter than those of previous studies at $z \sim 5$. A compilation of the 
quasar luminosity function at $4 \leq z \leq 6$ from the HSC-SSP suggests that there is little redshift 
evolution in the break magnitude and in the faint-end slope within this redshift range, although previous 
studies suggest that the faint-end slope becomes steeper at higher redshifts. The number density of 
low-luminosity quasars decreases more rapidly from $z \sim 5$ to $z \sim 6$ than from $z \sim 4$ to 
$z \sim 5$.
\end{abstract}


\section{Introduction} \label{sec:intro}

Active galactic nuclei (AGNs) release huge radiative energy, which is believed to be powered by the 
gravitational energy of the accreting medium to supermassive black hole (SMBH) at galactic centers 
\citep[e.g.,][]{ree84}. The mass of SMBHs ($M_{\rm BH}$) in quasars, the most luminous class of AGNs, 
reaches up to $\sim10^9 M_\odot$ or even higher \citep[e.g.,][]{wil10, she12}. Interestingly, such massive 
SMBHs are seen even at very high redshifts, $z \sim 6-7$ \citep[e.g.,][]{kur07, mor11, ven13, ven15, 
wu15, ban18, she19, ono19}. The previous studies suggest that the luminous quasars discovered at high 
redshift have large black hole mass ($M_{\rm BH}$) and accrete mass at the rate close 
to the Eddington limit. In the local Universe, it has been observationally revealed that 
there is a tight correlation between the mass of the host bulges ($M_{\rm bulge}$) and $M_{\rm BH}$ 
\citep[e.g.,][]{mahu03, hari04, gul09}. This correlation may suggest that SMBHs and their host galaxies 
have evolved together with close interplay that is now recognized as the galaxy-SMBH coevolution. 
Therefore, observational studies on the redshift evolution of SMBHs are important not only to understand 
the evolution of SMBHs, but also to understand the total picture of the galaxy evolution. 

The present work focuses on quasars. We refer to luminous, optically selected, unobscured (type-1) 
AGNs as quasars in our paper. Quasars are important, because (i) they are in the 
phase of rapid SMBH evolution via active gas accretion, and (ii) their huge luminosity 
enables us to 
estimate $M_{\rm BH}$ through spectroscopic observations, even in the very distant Universe. 
Measurements of the quasar luminosity function (QLF) and black hole mass function in a wide redshift 
range are a promising approach to reveal the cosmological evolution of SMBHs.

The QLF at $z \lesssim 4$ has been measured previously over a wide luminosity range \citep[e.g.,][]
{sia08, cro09, gli10, gli11, ike11, mas12, ros13, pal13, aki18, bou18, ste18, ada19}, leading to the 
recognition that the QLF is expressed by the broken power-law formula. At $z \gtrsim 5$, the brighter 
part of the QLF than knee of the LF has been measured \citep{ric06, she12, yan16, 
jia16, wil10}. While 
the faint side of the QLF at $z = 6$ has been established by the Subaru High-$z$ Exploration of 
Low-Luminosity Quasars project (SHELLQs; \citealt{mat16, mat18a, mat18b, mat18c}), the faint side of 
the QLF at $z \sim 5$ is still controversial \citep[e.g.,][]{ike12, mat13, mcg13, mcg18, nii16}. This is 
because the area and the depth of previous surveys are not enough to measure statistical properties of 
faint quasar at $z \sim 5$. 

However, the determination of the faint side of the QLF at $z \sim 5$ is important to 
understand the overall picture of SMBH growth in the early Universe.
Previous observations suggest that the number density of luminous quasars increased from the early 
Universe to $z \sim 2$ and then decreased to the current Universe \citep[e.g.,][]{ric06, cro09}. It is 
of particular interest to study possible luminosity dependences of the quasar 
number-density evolution (luminosity-dependent density evolution; LDDE). Recent optical surveys of 
high-redshift quasars have reported that low-luminosity quasars show the peak of the 
number density at lower redshifts than do high-luminosity quasars \citep[e.g.,][]{cro09, ike11, ike12, nii16}. 
Since the quasar luminosity at a given Eddington ratio corresponds to $M_{\rm BH}$, the reported LDDE 
trend is sometimes called the downsizing evolution. The same trend has been found
by X-ray surveys for all AGN populations including obscured (type-2) ones (e.g., \citealt{ued03, ued14, 
has05, miy15, air15}; see also \citealt{eno14, shi19} and references therein for theoretical works on the 
AGN downsizing evolution). Note that the downsizing evolution has been originally proposed to describe 
the redshift evolution of the galaxy mass function \citep[e.g.,][]{cow96, nei06, fon09}. Therefore the AGN 
downsizing evolution, if present, may provide a significant insight to the galaxy-SMBH co-evolution. 
However, the number density of low-luminosity quasars at high redshifts has been 
quite uncertain, which has prevented us from understanding the whole picture of the quasar LDDE. 
\citet{ued14} noted a possibility that the number ratio of total (type-1 and type-2) low-luminosity AGNs to 
high luminosity ones may increase from $z\sim3$ to $z\sim5$ (``up-down sizing''; see also \citealt{gli10, 
gli11, gia15}). We need large samples of high-$z$ low-luminosity AGNs to examine such scenarios.

The Subaru Strategic Program of the Subaru telescope with the Hyper Suprime-Cam (HSC-SSP; 
\citealt{aih18a}) provides an unprecedented dataset to search for low-luminosity quasars at high 
redshift. The survey started in March 2014 and is assigned 300 nights for 5 years. Taking advantage of 
the wide field-of-view (FoV) of HSC \citep[a circular FoV with 1.5 deg in diameter; see][]{miy18, kom18, 
kaw18, fur18}, the Wide-layer component of the HSC-SSP survey will cover 1,400 deg$^2$ when 
completed, mainly in the equatorial region. The $5\sigma$ survey depth of the Wide-layer component 
reaches down to 26.8, 26.4, 26.4, 25.5, and 24.7 AB mags in the $g$, $r$, $i$, $z$, 
and $y$ bands respectively for point sources, which are $\sim3$ mag deeper than the Sloan Digital Sky 
Survey (SDSS; \citealt{yor00}). 

The QLF at $z = 4$ and $z = 6$ have been derived with the HSC-SSP data by \citet{aki18} and 
\citet{mat18c}, respectively. In this paper, we derive the faint-end of the QLF at $z \sim 5$ with the 
HSC-SSP data. This paper is organized as follows. The selection criteria adopted in this work are 
described in Section \ref{sec:sample}. The survey completeness and contamination are described in 
Section \ref{sec:comple_contami}. The derived QLF is shown in Section \ref{sec:results}. In Section 
\ref{sec:discussion}, we compare the derived QLF parameters with those in the literature at $z > 3$ 
and discuss the evolution of the quasar number density. Finally, a summary is presented in Section 
\ref{sec:summary}. In the Appendix A, we show the results from our spectroscopic observations for 
a part of our sample of candidates. Throughout this paper, we adopt a $\Lambda$CDM cosmology 
with $\Omega_m = 0.3$, $\Omega_\Lambda = 0.7$, and the Hubble constant of $H_0 = 70$ 
km s$^{-1}$ Mpc$^{-1}$. All magnitudes are described in the AB magnitude system. We use 
PSF (point-source function) magnitudes for point sources and all the magnitudes are 
corrected for Galactic extinction based on dust maps by \citet{sch98}.

\section{Sample selection} \label{sec:sample}

\subsection{The HSC Data} \label{subsec:data}

We selected $z \sim 5$ quasar candidates from the HSC-SSP survey data. We used the S16A-Wide2 
internal release data from the Wide-layer component. The data were reduced with hscPipe-4.0.2 
\citep{bos18}. The S16A-Wide2 data cover an area of 339.8 deg$^2$ with all the five bands. However, 
the edge regions have shallower limiting magnitudes, so we used the full-depth region in the 5 bands 
in this work. The full-depth regions can be identified with the hscPipe parameter 
{\tt countinputs}, which records the number of exposures at a given position for each 
filter. For the Wide layer data, the full-depth regions correspond to {\tt countinputs} 
$\geq$ (4, 4, 6, 6, 6) for ($g$, $r$, $i$, $z$, $y$) \citep[e.g.,][]{aih18a}. In addition, we removed some 
problematic patches\footnote{
   ``Patch'' is a sub-region in HSC images, covering $\sim$$12 \times 12$ arcmin$^2$. The unit that 
   consists of $9 \times 9$ patches is called ``tract'' that covers $\sim$$1.7 \times 1.7$ deg$^2$.}. 
In some patches the color sequence of stars shows a significant offset from that computed from the 
Gunn-Stryker stellar spectrophotometric library \citep{gun83}, because hscPipe is unable to model the 
PSF accurately for the visits with an extremely good seeing \citep{aih18b}. The offset in each patch is 
given in the patch\_qa table stored in the HSC-SSP database. We removed patches, in which color 
offset is larger than 0.075 mag either in the $g - r$ vs. $r - i$, $r - i$ vs. $i - z$, or $i - z$ vs. $z - y$ 
color-color plane \citep[see Section 5.8.4 in][]{aih18a}. The Wide layer consists of 7 fields, i.e., 
GAMA09H, GAMA15H, WIDE12H, XMM-LSS, HECTMAP, VVDS, and AEGIS \citep{aih18a}. Since 
many patches in the VVDS and a part of the GAMA09H at Decl. $> 2$ deg are affected by the 
photometric offset problem, we removed all patches in these fields.  

Next we created a clean sample with reliable photometry using flags for the 5 bands provided by the 
hscPipe. First we selected objects detected in all of $r$, $i$, and $y$ bands. Then
we removed objects, which are affected by saturation, cosmic rays, or bad pixels in the 
central $3 \time 3$ pixels. We require that the sources are not blended and are in 
the inner region of a patch and a tract. We also require that an object is not close to bright sources. 
These flag criteria are listed in Table \ref{tab:tab_flag}. As a result, $\sim$30 million objects were 
selected as a clean sample in the area of 81.8 deg$^2$. This survey area was 
estimated by utilizing the HSC-SSP random catalog \citep{aih18b, cou18}, by 
counting the number of random points falling on the unflagged area. Thus the incompleteness of the 
flag selection criteria is taken into account (see Section~\ref{subsec:completeness} for general 
treatments of the survey completeness).

Since high-redshift quasars should be observed as point sources, we selected only point sources from 
the clean sample. For this purpose, we used the second order adaptive moments. We used the 
adaptive moments measured in the $i$-band, since $i$-band images were preferentially taken under 
good seeing conditions ($\sim$0.58$\pm$0.08 arcsec in the Wide layer\footnote{
   This is slightly larger than the value that \citet{aih18b} reported in their paper (0.56
   arcsec). This is because \citet{aih18b} reported the median seeing size for the entire Wide-layer 
   regions while we are focusing only on the full-depth and full-color regions.
}). 
The adaptive moment is measured based on the algorithm of \citet{hir03}. Our point 
source criteria are defined with the ratios of the second order adaptive moments to those of the PSF 
at the position of a given object: 
\begin{eqnarray}
\rm{ishape\_hsm\_moment\_11} &/& \rm{ishape\_hsm\_psfmoment\_11} \nonumber \\ &<& 1.1,\\
\rm{ishape\_hsm\_moment\_22} &/& \rm{ishape\_hsm\_psfmoment\_22} \nonumber \\ &<& 1.1 \label{equ2},
\end{eqnarray}
where the indices of ``11'' and ``22'' denote the East-West direction and 
North-South direction, respectively. The completeness and contamination inherent to this selection were evaluated in \citet{aki18} by comparing the 
HSC $i$-band measurements with those by the Hubble Space Telescope (HST) Advanced Camera 
for Surveys (ACS) (see Section 2.2 in \citealt{aki18} and Section \ref{subsec:contamination} in this 
paper for more details). Since (i) the point source selection becomes uncertain for
faint objects and (ii) bright objects are saturated, we focus only on objects with $19.1 < i < 24.1$ 
where the incompleteness and contamination are low. The above procedure 
selected 968,997 objects. 

\begin{deluxetable}{lc}
\tablecaption{Flag values used to construct the clean sample\label{tab:tab_flag}}
\tablehead{
\colhead{Flag} & \colhead{condition}
}
\startdata
flags\_pixel\_saturated\_center & False \\
flags\_pixel\_cr\_center & False \\
flags\_pixel\_bad & False \\
flags\_pixel\_edge & False \\
detect\_is\_primary & True \\
flags\_pixel\_bright\_object\_any & False \\
\enddata
\end{deluxetable}

\subsection{Color selection} \label{subsec:color}

\begin{figure}[ht!]
\includegraphics[width=9cm]{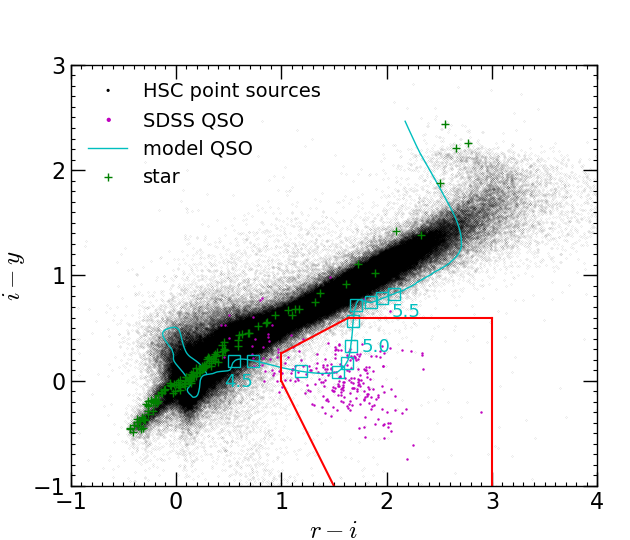}
\caption{
   Two-color ($i - y$ vs. $r - i$) diagram of the HSC point sources with $19.1 < i < 24.1$ (black dots) 
   and model stars of \citet{pic98} (green crosses). The color track of model quasars described in 
   \citet{nii16} is shown as a cyan line. The cyan squares along the cyan line represent redshifts from 
   4.5 to 5.5 in steps of 0.1. The magenta dots represent the SDSS quasars at $4.4 \leq z < 5.6$. 
   Our quasar selection criteria are shown with the red lines.
\label{fig:fig_riy}}
\end{figure}

We selected $z \sim 5$ quasar candidates by colors with the following two steps. First we used the 
two-color diagram of $i - y$ vs. $r - i$. The colors of the HSC point sources selected in Section 
\ref{subsec:data} are shown in Figure \ref{fig:fig_riy}. Also shown are the colors of stars, which are 
calculated from stellar spectra of \citet{pic98} with the HSC filter response curves \citep{miy18}. The 
color sequence of the HSC point sources is roughly consistent with that of library stars. We show the 
color track of a model quasar in Figure \ref{fig:fig_riy}. This model corresponds to the average 
spectrum of quasars with $M_i [z = 2] = -26.5$, which was derived in \citet{nii16} (see also Section \ref{subsec:completeness}). We also show the colors of SDSS DR12 quasars at 
$4.4 \leq z < 5.6$ (324 objects, selected by the following criteria; $4.4 \leq z_{\rm pipe} \leq 5.6$, 
BAL\_FLAG = 0, ZWARNING = 0, $\abs{\rm{err\_zpipe}}$ $<$ 0.001, $\abs{z_{\rm vi} - z_{\rm pipe}}$ $<$ 0.05, 
SNR\_SPEC $>$ 1, S/N $>$ 3 for $r_{\rm HSC}$, and S/N $>$ 5 for $i_{\rm HSC}$ and 
$y_{\rm HSC}$; see \citealt{par17}), 
whose HSC magnitudes were estimated from the SDSS magnitudes by using Equations (9), (10), 
and (12) in \citet{aki18}. Our selection criteria of $z \sim 5$ quasar candidates from the HSC 
point-source sample are: 
\begin{eqnarray}
   0.53 (r - i) - 0.27 &>& (i - y), \label{eq:criteria_riy1}\\ 
   -2.0 (r - i) + 2.0 &<& (i - y), \label{eq:criteria_riy2}\\ 
   1.0 < (r - i) &<& 3.0, \\ 
   (i - y) &<& 0.6. \label{eq:criteria_riy4}
\end{eqnarray}
The criterion of Equation (\ref{eq:criteria_riy1}) was determined as follows. We fitted the color of HSC 
point sources with a linear function in the two-color plane. In this process, we only used point sources 
that satisfy Equation (\ref{eq:criteria_riy2}) and $(r - i) > 1.0$. To investigate the distribution of the color 
of galactic stars with removing contamination of point-like galaxies, we imposed 
Equation~(\ref{eq:criteria_riy2}). Then we calculated their standard deviation and determined the 
criterion at the 3$\sigma$ separation from the best fit, which corresponds to 
Equation~(\ref{eq:criteria_riy1}). In addition to this Equation~(\ref{eq:criteria_riy1}), we also adopted
Equations~(\ref{eq:criteria_riy2})--(\ref{eq:criteria_riy4}) to avoid contaminations by Galactic stars.
By these criteria, 613 objects were selected.

\begin{figure}[ht!]
\includegraphics[width=8.8cm]{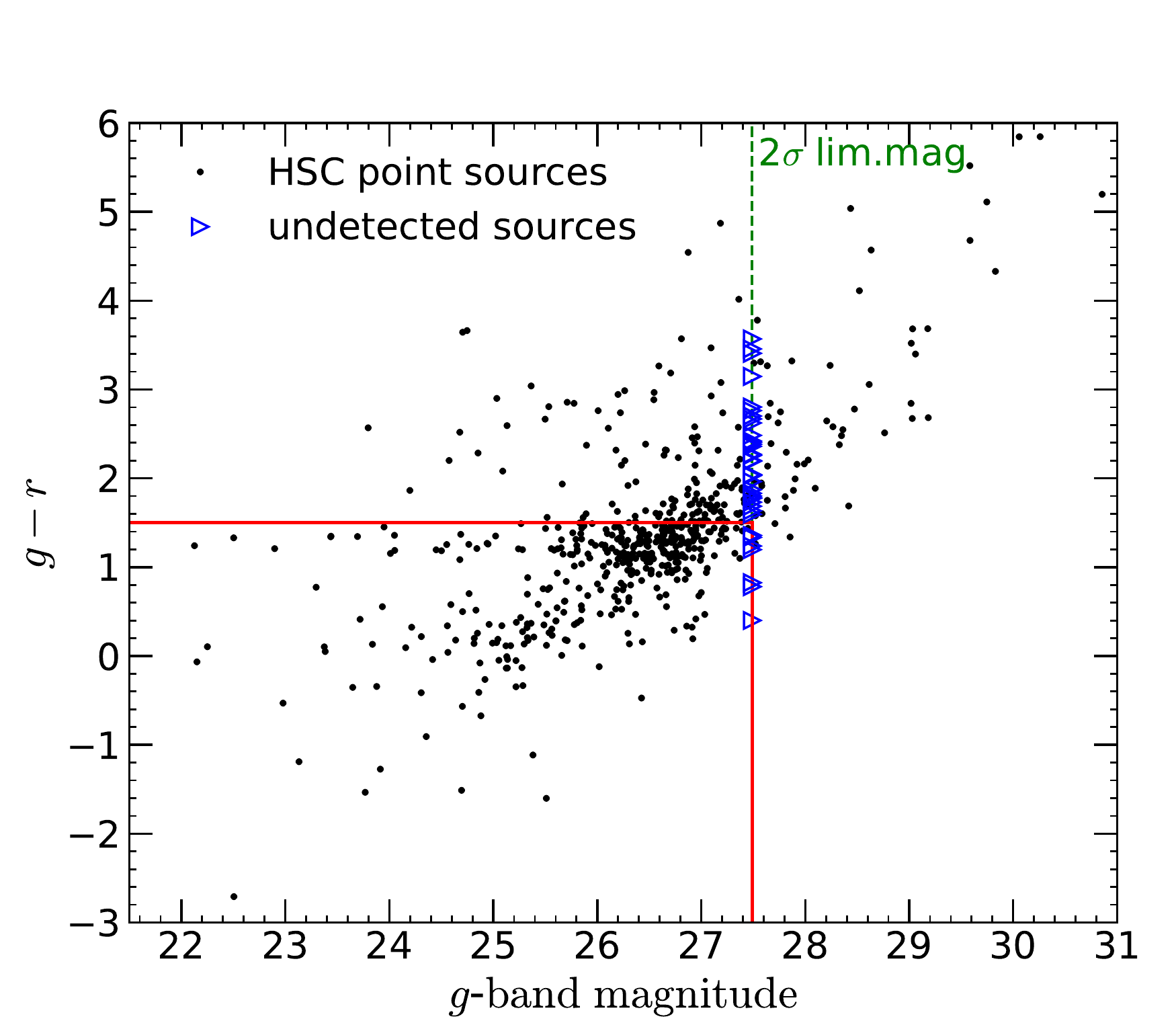}
\caption{
   Color selection with a $g - r$ vs. $g$ diagram. The HSC point sources selected by Equations 
   (\ref{eq:criteria_riy1}) - (\ref{eq:criteria_riy4}) are represented by the black dots. Point sources 
   undetected in the $g$-band are shown by the blue triangles placed at $g=24.79$. 
   The green dashed line shows the $g$-band $2\sigma$ limiting magnitude. Our selection criteria are 
   shown by the red lines; the point sources in the upper or right side of the red lines 
   are selected. 
\label{fig:fig_gr}}
\end{figure}

In addition, we used $g - r$ color to further remove the remaining sources of contamination,
especially at lower redshift. Quasars at $z \sim 5$ have red $g - r$ colors due to the IGM absorption. 
Based on the color distribution of model quasars, we determined the criterion,
\begin{eqnarray}
   (g - r) > 1.5, \label{eq:criteria_gr}\\
   \rm{or} \ \ \ \ \ \ \nonumber \\ 
   g \geq 27.49, \label{eq:criteria_g}
\end{eqnarray}
where Equation~(\ref{eq:criteria_gr}) was determined by checking the color of 
quasars (see, e.g., Figure~4 in \citealt{ike12}).
Figure \ref{fig:fig_gr} shows a diagram of $g - r$ colors versus $g$-band magnitudes of the HSC point 
sources selected by Equations (\ref{eq:criteria_riy1}) -- (\ref{eq:criteria_riy4}). For $g$-undetected 
objects, we assigned the typical 2$\sigma$ limiting magnitude ($g = 27.49$) in this figure, 
which satisfies Equation~(\ref{eq:criteria_g}).
By these criteria we selected 241 objects. 

Finally we removed 17 objects by visual inspection. Those 17 
objects appear to be significantly affected by artificial features that are related to cosmic rays, 
tails of bright stars, and so on, which were not perfectly removed by the flags 
in Table \ref{tab:tab_flag}. Thus the final candidates of $z \sim 5$ quasars are 
224 objects. The sample of our HSC candidates contain no previously-known, 
spectroscopically-confirmed quasars.

\section{Survey completeness and contamination} \label{sec:comple_contami}

\subsection{Completeness} \label{subsec:completeness}

In order to assess the number density of quasars at $z \sim 5$, the survey 
completeness should be evaluated. We estimated the detection completeness and selection
completeness as functions of the $i$-band apparent magnitude and the quasar redshift, and then 
evaluated the survey completeness by combining those two factors. Here the detection completeness 
is the fraction of quasars that are detected in all of $r$, $i$, and $y$ bands in the observed images,
while the selection completeness is the fraction of quasars whose color properties with photometric 
errors satisfy the selection criteria. 

For evaluating the completeness, we constructed quasar models as follows \citep[see also][]{nii16, 
aki18}. We assumed that the spectral energy distribution (SED) of quasars is independent of the 
redshift \citep[see, e.g.,][]{kuh01, tel02, yip04, jia06, nii16} and also the luminosity \citep{tel02}. 
We adopted a double power-law continuum ($f_\nu \propto \nu^{-\alpha_\nu}$), with a spectral break 
at $\lambda_{\rm{rest}} = 1100$ \AA. The adopted spectral slope at the shorter wavelength side is 
$\alpha_\nu = 1.76$ \citep{tel02} while that at longer wavelength side is $\alpha_\nu = 0.46$ \citep{van01}. 
The standard deviation of the slope is assumed to be 0.30 at the both sides \citep{fra96}. Strong 
emission lines were added to this continuum, assuming Gaussian profiles. Here we included emission 
lines whose flux is larger than 0.5\% of the Ly$\alpha$ flux. Since EW of broad emission lines depends 
on the quasar luminosity in the sense that lower-luminosity quasars have a larger EW \citep[the Baldwin 
effect;][]{bal77, kin90, bas04, nag06}, we adopted the luminosity-dependent EW$_{\rm rest}$(C~{\sc iv}) 
taken from Table 2 of \citet{nii16} and flux ratios given in Table 2 of \citet{van01}. The scatter of 
EW$_{\rm rest}$(C~{\sc iv}) is taken into account in the spectral modeling. We also included the Balmer 
continuum and the Fe~{\sc ii} multiplet emission features by adopting the template given by \citet{kaw96}. 
The effects of the intergalactic absorption by neutral hydrogen were incorporated by adopting the 
opacity models of \citet{ino14}, considering the scatter of the hydrogen column density estimated with a 
Monte Carlo method described in \citet{ino08}. We constructed 1,000 quasar SED models in each 
$i$-band magnitude and redshift bin ($\Delta i = 0.25$ mag and $\Delta z = 0.1$) spanning 
$19.1 \leq i \leq 24.1$ mag and $4 \leq z \leq 6$. In total 441,000 model quasar SEDs were constructed.

\begin{figure}[ht!]
\includegraphics[width=8.55cm]{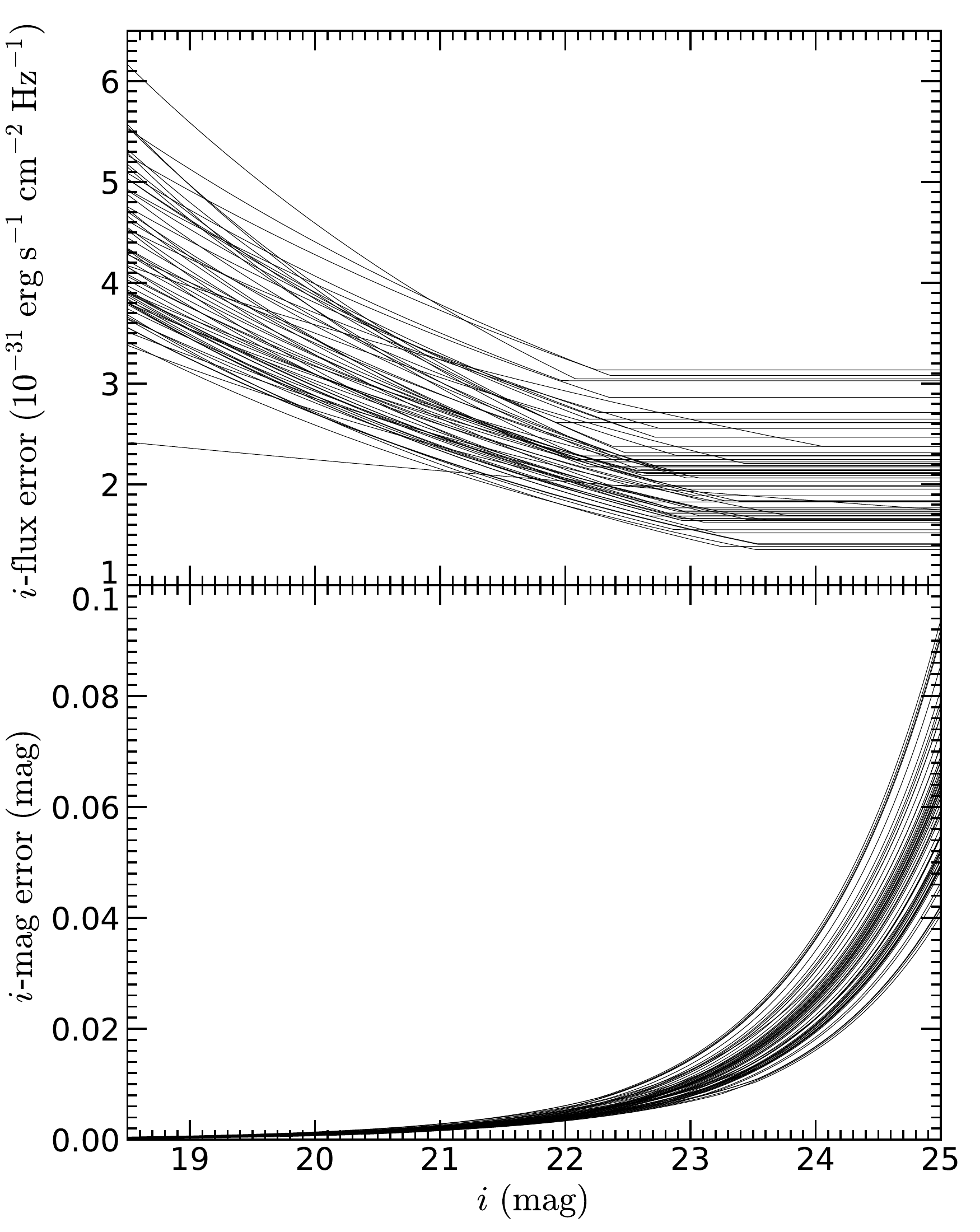}
\caption{
   Results of the fit to the photometric error of the HSC point sources for $i$-band in the randomly selected 
   62 patches. The errors are shown as flux density at the top panel and magnitude at the bottom panel.
\label{fig:fig_flux_error}}
\end{figure}

Then we added the photometric error to each magnitude of the model quasars. We investigated the flux 
errors of real HSC point sources in each of the $g$, $r$, $i$, and $y$ bands on randomly selected 62 
patches. We fitted a linear function to those flux errors in the logarithmic scale as a function of magnitude 
in the brighter side (which corresponds to the photon noise), and calculated the average of the flux error 
in the fainter side (which corresponds to the sky noise). In this calculation we applied $3\sigma$ clipping 
in each magnitude bin ($\Delta$mag = 0.2). The results of the fit to the $i$-band photometric error of 
point sources in each of 62 patches are shown in Figure \ref{fig:fig_flux_error}. For each model quasar, 
photometric error was assigned in each band assuming Gaussian error distribution with the standard 
deviation estimated above. For each model quasar, we generated 10,000 realizations with different 
amounts of additional noise.

\begin{figure}[ht!]
\includegraphics[width=8.55cm]{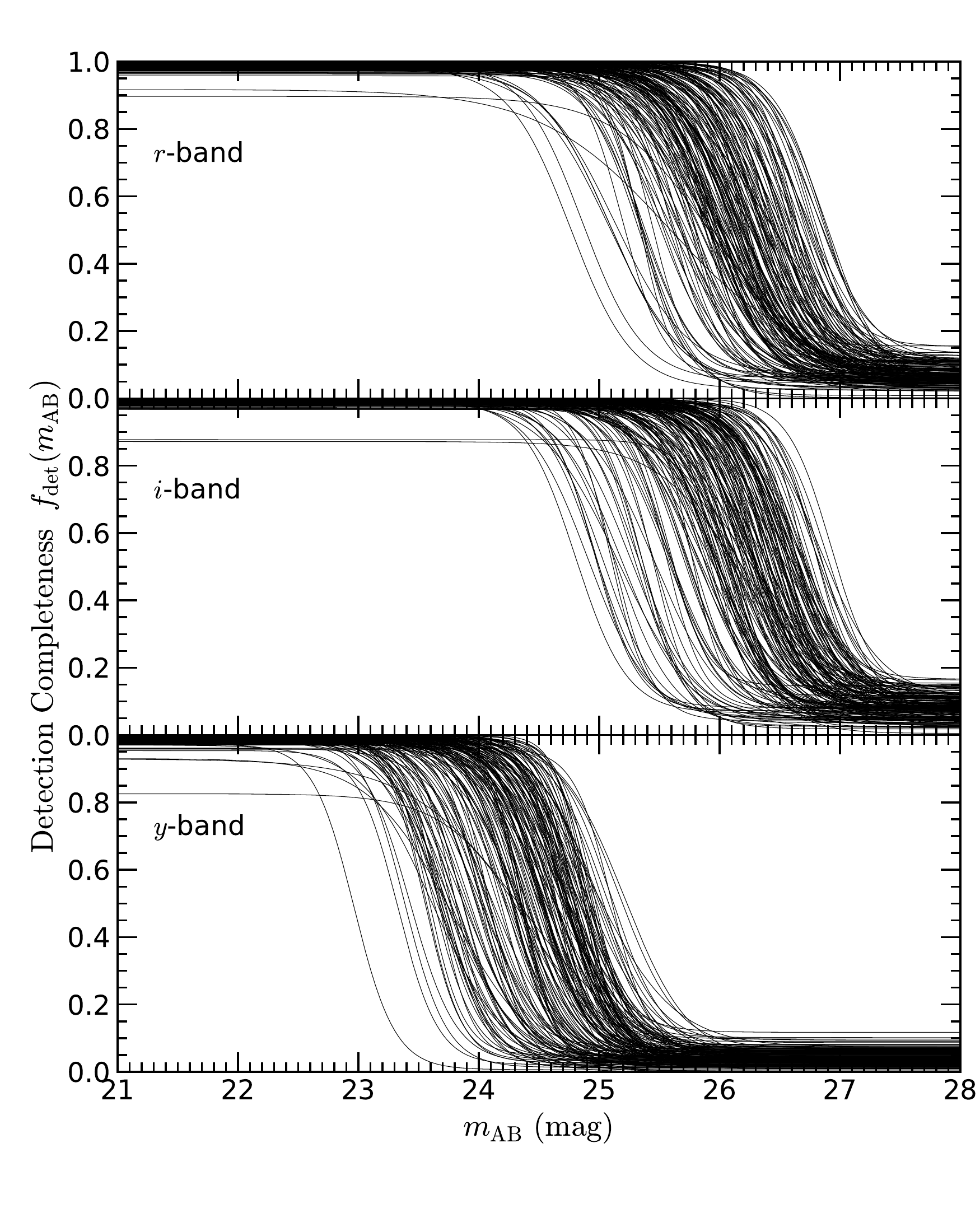}
\caption{
   Detection completeness in the $r$ (top), $i$ (middle), and $y$ (bottom) bands as modeled by Equation
   \ref{eq:detection_completeness}, measured in the randomly selected 62 patches. 
\label{fig:fig_detec_compl}}
\end{figure}

To estimate the detection completeness of quasars, we first investigated the detection completeness of
point sources in each of the $r, i$, and $y$ bands. We adopted the same method as described in 
\citet{aih18b}, i.e., artificial point sources were inserted at random positions of the stacked HSC images, 
and tried to recover them with hscPipe. We ran the simulations for point sources with 
$19.1 \leq i \leq 24.1$ mag in the randomly selected 62 patches as described above. Then we fitted the 
derived detection completeness as a function of magnitude in each band with a function of \citep{ser00}:
\begin{eqnarray}
   f_{\rm det} (m_{\rm AB}) = \ \ \ \ \ \ \ \ \ \ \ \ \ \ \ \ \ \ \ \ \ \ \ \ \ \ \ \ \ \ \ \ \ \ \ \ \ \ \ \ \ \ \ \ \ \ \ \nonumber \\
   \frac{(f_{\rm{max}} - f_{\rm{min}})}{2} (\tanh [\alpha (m_{50} - m)] + 1) + f_{\rm{min}}, 
   \label{eq:detection_completeness}
\end{eqnarray}
where $f_{\rm{max}}$, $f_{\rm{min}}$, $\alpha$, and $m_{50}$ represent the detection completeness at 
the brightest and faintest magnitudes, the sharpness of the transition between $f_{\rm{max}}$ and 
$f_{\rm{min}}$, and the magnitude where the detection completeness is 50\%, respectively. In most 
cases, the completeness at the faintest magnitudes ($f_{\rm{min}}$) is higher than zero, which is likely 
due to chance coincidences of input sources with real brighter sources that exist in HSC images. 
The fitting results for the 62 random patches are shown in Figure \ref{fig:fig_detec_compl}.  
In the following, we fixed parameters $f_{\rm{max}}$ and $f_{\rm{min}}$ to 1.0 and 0.0, respectively.

\begin{figure}[ht!]
\includegraphics[width=8.8cm]{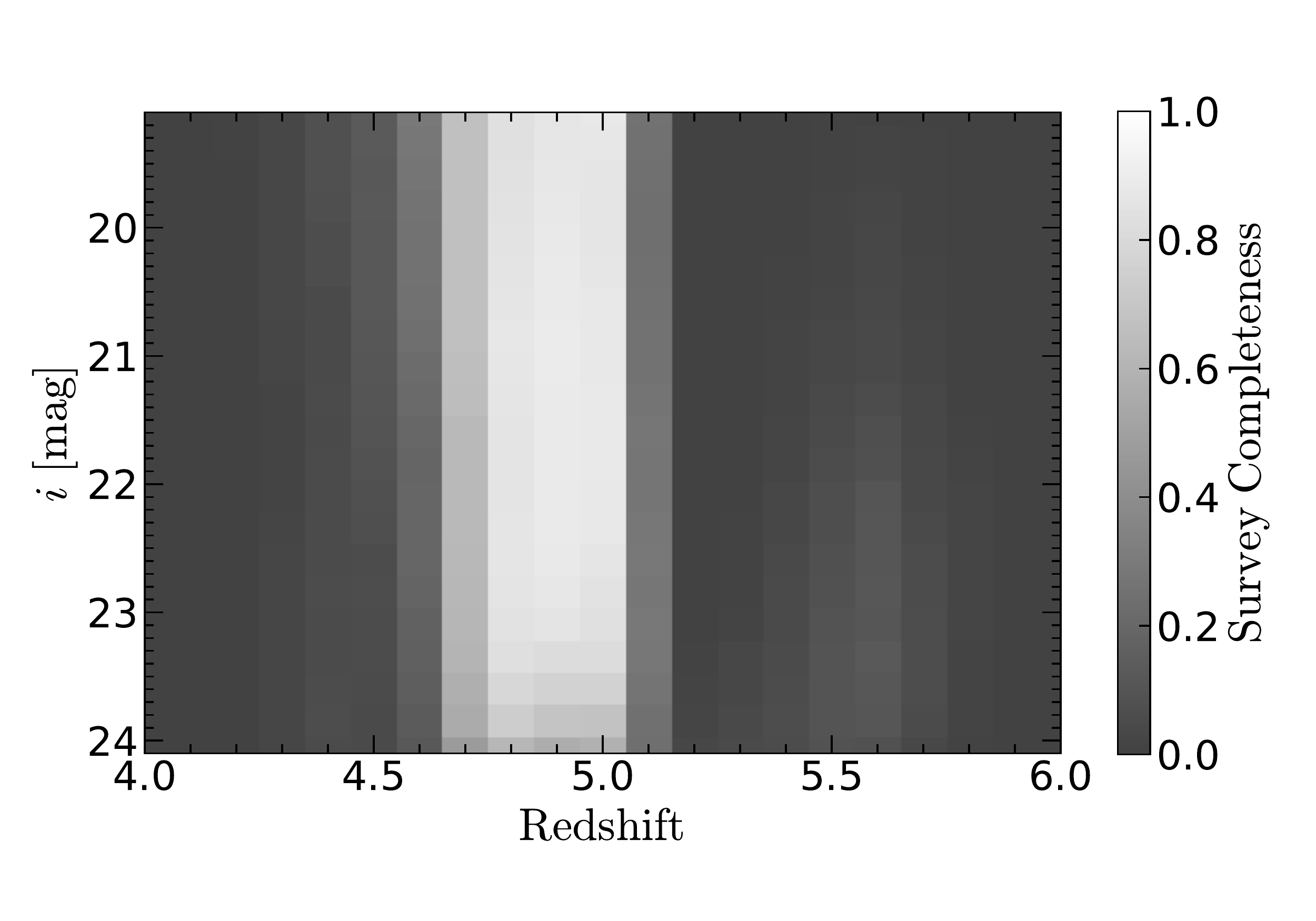}
\caption{
   Survey completeness as a function of $i$-band magnitude and redshift, ranging from 1.0 in white to 
   0.0 in dark gray.
\label{fig:fig_completeness}}
\end{figure}

To evaluate the probability that a quasar is detected in all of the $r$, $i$, and $y$-band images, we 
multiplied the detection probability of each simulated quasar in $r$, $i$, and $y$-band images. Then we
calculated the average of this result for 1,000 model quasars with a certain $i$-band magnitude and
redshift. By calculating this average for all model quasars in the ranges of $19.1 \leq i \leq 24.1$ mag 
and $4 \leq z \leq 6$, we obtained the detection completeness of quasars as a function of $i$-band 
magnitude and redshift.
The selection completeness is also estimated as a function of $i$-band magnitude and redshift, by
calculating the fraction of model quasars that satisfy our color selection criteria.
Finally, by multiplying the detection completeness and selection completeness of quasars,
we evaluated the survey completeness as a function of $i$-band magnitude and redshift. This survey 
completeness is shown in Figure \ref{fig:fig_completeness}. This figure suggests that 
the survey completeness of our quasar survey is high; the averaged completeness at 
$4.7 < z < 5.1$ and $i < 23.6$ mag is 0.87.

Note that the survey completeness shown in Figure~\ref{fig:fig_completeness} does not include the
completeness for the point-source criteria (Equations 1 and 2). As described in \citet{aki18}, the 
completeness of the point-source criteria is higher than 90\% for objects with $i<23$ 
while it drops to $\sim$70\% for objects with $i=24$, when we refer 
the ``best seeing conditions'' in \citet{aki18} (note that the $i$-band images of the HSC-SSP Wide 
layer were obtained with the seeing size of 0.58$\pm$0.08 arcsec, which is close to the
``best seeing conditions (0.5 arcsec)'' rather than the ``median seeing conditions (0.7 arcsec)'' in
\citealt{aki18}; see Section~\ref{subsec:data}). We did not correct for this incompleteness because 
we fit the QLF only in the range of $i<23.1$ (see Section~\ref{subsec:Double_power-law_model}).

\subsection{Contamination} \label{subsec:contamination}

\begin{figure}[ht!]
\includegraphics[width=8.8cm]{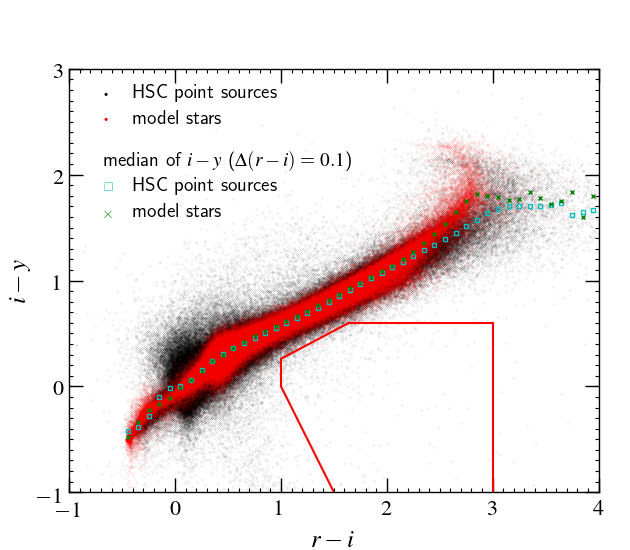}
\caption{The color distribution of the model stars (red dots) and the HSC point sources at $19.1 < i < 24.1$ (black dots) in the 
$i - y$ vs. $r - i$ two-color diagram. The median $i - y$ values in the $r - i$ bins ($\Delta (r - i) = 0.1$) are plotted with the cyan 
squares and the green crosses for the HSC point sources and the model stars, respectively.  
\label{fig:fig_modelstar_color}}
\end{figure}

Here we estimate the contamination rate in our photometric sample of $z \sim 5$ quasars. In order to 
investigate the contamination of Galactic stars, we simulated the number of model stars, which satisfy 
our color selection criteria. The model stars were constructed by TRILEGAL code, which simulates the 
number and magnitudes of stars in a given field \citep{gir05}. We adopted an exponential disk model 
with the default values of scale length and height, and the Chabrier initial mass 
function \citep{cha03}. 
As a result, we got 810,178 model stars. We converted the output Pan-STARRS1 
\citep[PS1;][]{cha16} photometry to the HSC photometry by adopting the color terms, using the 
coefficients given in Table \ref{tab:color_term}; for example,
\begin{eqnarray}
   g_{\rm{HSC}} = g_{\rm{PS1}} + a0 + a1 \times (g - r)_{\rm{PS1}} + a2 \times (g-r)_{\rm{PS1}}^2 
   \nonumber \\     
\end{eqnarray}
(\citealt{aih18b}; the web page of the HSC-SSP public data release 1 
\footnote{\url{https://hsc-release.mtk.nao.ac.jp/doc/index.php/data/\#color-terms}}). We added error to 
the HSC photometry of the model stars in the same way as we did for the model quasars (Section 
\ref{subsec:completeness}). The color distribution of the model stars ($19.1 < i < 24.1$) in the $i - y$ vs. 
$r - i$ two-color diagram is shown in Figure~\ref{fig:fig_modelstar_color}. The color of the real HSC point 
sources in the same magnitude range is also shown in the figure. 
The median of the observed and model $i - y$ colors are almost consistent with each 
other in each $r - i$ bin with $\Delta (r - i) = 0.1$, as shown in Figure~\ref{fig:fig_modelstar_color}. 
The color scatter of the HSC point sources is somewhat larger than that of the model stars, which 
could be possibly due to the underestimate of the HSC photometric error measured by the HSC pipeline
as reported in \citet{aih18b}. This may lead to the underestimate of the contamination rate, which should
be kept in mind in the later analysis. But here we adopt the models of Galactic stars shown in 
Figure~\ref{fig:fig_modelstar_color}. In this case, the number of model stars that meet our color selection 
criteria is 14 at $23.6 < i < 24.1$, while no model stars at $i < 23.6$ satisfy our color selection criteria.
Therefore the contamination rate is negligibly small at a brighter range ($i < 23.6$) and moderately low 
at a fainter range (14/114 $\sim$12\% at $23.6 < i < 24.1$).

Note that, some HSC point sources have photometric errors that are not due to the Gaussian 
background fluctuation, e.g., cosmic rays. Based on our individual checks, we found 
that such objects distribute at a distance from the 
color sequence of Galactic stars in the $i-y$ vs. $r-i$ two-color diagram and partly enter our selection 
region for $z \sim 5$ quasar candidates. Since such objects are mostly faint ($i > 23.1$) objects, our 
quasar candidates at $i > 23.1$ may include such objects as contaminations. In addition, as shown in 
Figure 1 of \citet{aki18}, the contamination rate of compact galaxies that is not spatially resolved in 
HSC images is almost 0\% at $i < 23.1$, while it increases from 0 to $\sim$35\% toward $i = 24.1$. 
The contamination of such compact galaxies is partly the origin of the difference between 
the number of HSC point sources (968,997) and the number of model stars (810,178). Compact 
galaxies in the HSC point-source sample also cause a larger scatter of colors in two-color diagram 
than the color scatter of model stars seen in Figure \ref{fig:fig_modelstar_color}. For the QLF 
fitting, we excluded the data at $i > 23.1$ as described later in Section \ref{subsec:binnedQLF}.
 
For four objects with $i < 23.2$ in our photometric sample of quasars at $z \sim 5$, we 
conducted spectroscopic observations. The observations were carried out with the 4m Blanco telescope 
at the Cerro Tololo Inter-American Observatory (CTIO) and the Subaru Telescope at the Hawaii 
Observatory of National Astronomical Observatory of Japan (NAOJ). The details of the spectroscopic 
observations are described in Appendix \ref{sec:spec}. All of the four observed 
candidates are confirmed as $z \sim 5$ quasars. This result is consistent with the above estimate that 
the contamination rate is low, but it should be kept in mind that those four objects are biased toward the 
brightest objects in our sample of candidates.

\begin{deluxetable}{ccrrr}
\tablecaption{The parameters of the color term\tablenotemark{a} \label{tab:color_term}}
\tablehead{
\colhead{HSC} & \colhead{PS1} & \colhead{a0} & \colhead{a1} & \colhead{a2}
}
\startdata
$g$ & $g - r$ & $0.00730066$ & $0.06508481$ & $-0.01510570$ \\
$r$ & $r - i$ & $0.00279757$ & $0.02093734$ & $-0.01877566$ \\
$i$ & $i - z$ & $0.00166891$ & $-0.13944659$ & $-0.03034094$ \\
$y$ & $y - z$ & $-0.00156858$ & $0.14747401$ & $0.02880125$ \\
\enddata
\tablenotetext{a}{
    See \citet{aih18b} for more details.}
\end{deluxetable}

\section{Results} \label{sec:results}

\subsection{Binned quasar luminosity function} \label{subsec:binnedQLF}

The numbers of our $z \sim 5$ quasar candidates are summarized in Table \ref{tab:tab_num_candidate}. 
For the HSC sample, we assume $z = 4.9$ to infer $M_{1450}$ because it corresponds 
to the peak of the HSC survey completeness (see Figure \ref{fig:fig_completeness}).
The quantities given in this 
table are corrected for the contamination of Galactic stars, and the following analysis is based on these 
corrected values. The effective comoving volume of our HSC survey is calculated as
\begin{eqnarray}
   V_{\rm {eff}}(m_{i}) = d\Omega \int_{z = 0}^{z = \infty} C(m_{i}, z) \frac{dV}{dz}dz,
\end{eqnarray}
where $d\Omega$ (81.8 deg$^2$) is the solid angle of our survey and $C(m_{i}, z)$ is the survey 
completeness estimated in Section~\ref{subsec:completeness}. We calculate the number density of the 
quasars as 
\begin{eqnarray}
   \Phi(m_{i}, z) 
     = \sum_{\rm{j}}\frac{1}{V_{\rm{eff}}^{\rm{j}}\Delta m_i} 
     = \frac{N_{\rm{cor}}}{V_{\rm{eff}} (m_i) \Delta m_i}, 
\end{eqnarray} 
where $\Delta m_{i} = 0.5$, and $N_{\rm{corr}}$ is the corrected number of quasars. The uncertainty of 
the number density is estimated with the Poisson statistics. In the magnitude bins where the corrected 
number of quasars is lower than 50, we calculate the uncertainty using the statistics presented in 
\citet{geh86}.

\begin{figure}[ht!]
\includegraphics[width=8.9cm]{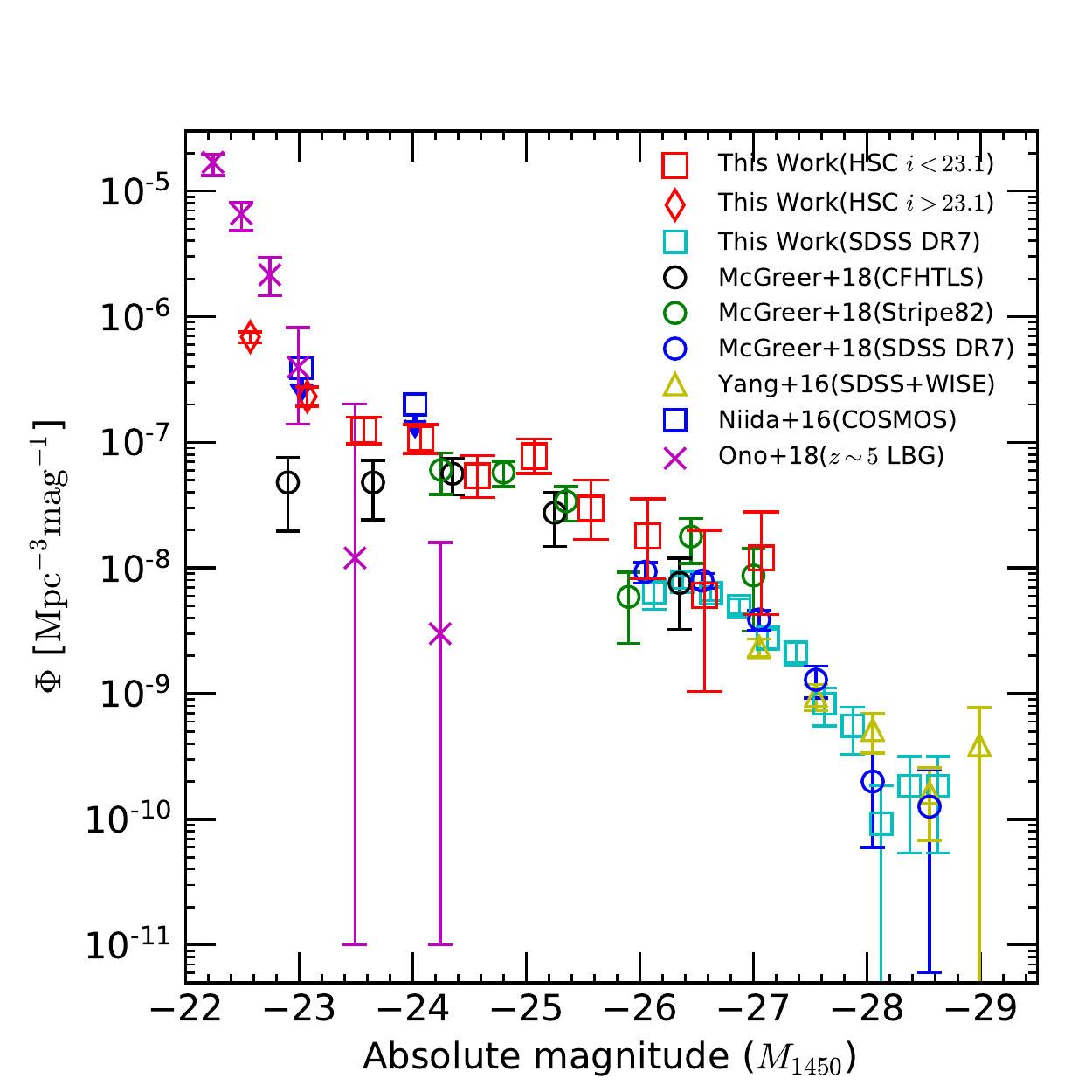}
\caption{
   The binned QLF at $z \sim 5$ derived in this work (red squares: HSC at $i < 23.1$, red diamonds: HSC 
   at $i > 23.1$, cyan squares: SDSS DR7). The black, green, and blue circles represent the $z = 4.9$ 
   QLFs of \citet{mcg18} measured with the data from CFHTLS, Stripe82, and SDSS DR7, respectively. 
   The yellow triangles are the quasar number densities at $z = 4.9$, which are converted from $z = 5.1$ 
   by using the redshift evolution given by \citet{fan01}, from SDSS+WISE survey \citep{yan16}. The blue 
   squares show the upper limit of the quasar number density at $z \sim 5$ derived from the COSMOS 
   quasar survey \citep{nii16}. For comparison we also show the LBG luminosity function at $z \sim 5$ 
   from the HSC-SSP survey \citep{ono18} with the magenta crosses.
\label{fig:fig_number_density}}
\end{figure}

The calculated number densities of quasars and their uncertainties are listed in Table 
\ref{tab:tab_num_candidate} and in Figure \ref{fig:fig_number_density}. Figure \ref{fig:fig_number_density} 
also shows the results from previous studies \citep{mcg18, yan16, nii16}. Our number densities at 
$M_{1450} < -24.32$ are consistent with those reported in the previous studies. At $M_{1450} > -23.3$ 
the number densities of Lyman break galaxies \citep{ono18} exceed that of quasars. In this luminosity 
range, the contamination rate of point-like compact galaxies increases from 0 to 35 \% as described in 
Section \ref{subsec:contamination} \citep[see Figure 1 in][]{aki18}. Therefore our quasar number densities 
are affected by such point-like galaxies at low-luminosity, and so we exclude the data at 
$M_{1450} > -23.32$ from the QLF fitting we describe below.

In order to constrain the bright-end of the QLF, we construct a high-luminosity quasar sample using the 
spectroscopically-confirmed sample of the SDSS DR7 quasar catalog \citep{sch10}. We select quasars 
at $z = 4.5 - 5.2$ from the catalog by following the recipe described in Section 4.3 of \citet{aki18}. 
We select those quasars with SCIENCEPRIMARY $= 1$ and “selected with the final quasar algorithm'' 
$= 1$ based on the TARGET photometry \citep[see Table 1 of][]{sch10}. The effective survey area of the 
SDSS DR7 sample is 6248 $\rm{deg}^2$ \citep{she12}. The selection completeness is estimated in 
\citet{ric06} as a function of redshift and magnitude. At $z > 4.5$ the selection efficiency is evaluated to 
be close to $\sim$$100$\% at $i < 20.2$ mag \citep[Figure 6 of][]{ric06}. At a given redshift, $z$, we select 
quasars with $M_i (z = 2) < -0.286 (z - 4.5) - 27.6$ as a complete sample with $100$\% completeness 
\citep[see Figure 17 in][]{ric06}. We convert $M_i (z = 2)$ to $M_{1450}$ with 
$M_{1450} = M_i (z = 2) + 1.486$ \citep[appendix B of][]{ros13} and calculate the effective survey volume. 
The evaluated number densities are shown in Figure \ref{fig:fig_number_density} and summarized in 
Table \ref{tab:tab_num_candidate}. The derived number densities are consistent with those reported in 
previous studies \citep{mcg18, yan16}.

\begin{deluxetable}{cccc}
\tablecaption{The number of the $z \sim 5$ quasar candidates\label{tab:tab_num_candidate}}
\tablehead{
\colhead{$M_{1450}$} & \colhead{$m_i$} & \colhead{$N_{\rm{corr}}$\tablenotemark{a}} & \colhead{$\it{\Phi}$\tablenotemark{b}} \\
\colhead{(mag)} & \colhead{(mag)} & & \colhead{(10$^{-8}$ Mpc$^{-3}$ mag$^{-1}$)}
}
\startdata
HSC S16A-Wide2 \\ \hline
$-22.57\tablenotemark{c}$ & 23.85 & 100 & $68.9^{+6.9}_{-6.9}$\\
$-23.07\tablenotemark{c}$ & 23.35 & 38 & $23.1^{+4.4}_{-3.7}$\\
$-23.57$ & 22.85 & 21 & $12.5^{+3.4}_{-2.7}$\\
$-24.07$ & 22.35 & 18 & $10.7^{+3.2}_{-2.5}$\\
$-24.57$ & 21.85 & 9 & $5.39^{+2.46}_{-1.76}$\\
$-25.07$ & 21.35 & 13 & $7.78^{+2.81}_{-2.13}$\\
$-25.57$ & 20.85 & 5 & $2.98^{+2.01}_{-1.29}$\\
$-26.07$ & 20.35 & 3 & $1.80^{+1.75}_{-0.98}$\\
$-26.57$ & 19.85 & 1 & $0.60^{+1.39}_{-0.50}$\\
$-27.07$ & 19.35 & 2 & $1.20^{+1.58}_{-0.78}$\\ \hline
SDSS DR7 \\ \hline
$-26.13$ & 20.30 &14 & 0.639$\pm$0.171\\
$-26.38$ & 20.05 & 79 & 0.772$\pm$0.087\\
$-26.63$ & 19.80 & 68 & 0.628$\pm$0.076\\
$-26.88$ & 19.55 & 54 & 0.499$\pm$0.068\\
$-27.13$ & 19.30 & 30 & 0.277$\pm$0.051\\
$-27.38$ & 19.05 & 23 & 0.212$\pm$0.044\\
$-27.63$ & 18.80 & 9 & 0.083$\pm$0.028\\
$-27.88$ & 18.55 & 6 & 0.055$\pm$0.023\\
$-28.13$ & 18.30 & 1 & 0.0092$\pm$0.0092\\
$-28.38$ & 18.05 & 2 & 0.018$\pm$0.013\\
$-28.63$ & 17.80 & 2 & 0.018$\pm$0.013\\
\enddata
\tablenotetext{a}{The numbers of candidates after subtracting the estimated stellar contamination.}
\tablenotetext{b}{The number densities of candidates derived with $N_{\rm{corr}}$.}
\tablenotetext{c}{Since the quasar number densities at $M_{1450} > -23.32$ may be affected by point-like 
galaxies, we exclude the data from the QLF fitting described in Section \ref{subsec:Double_power-law_model}.}
\end{deluxetable}

\begin{figure}[ht!]
\includegraphics[width=9cm]{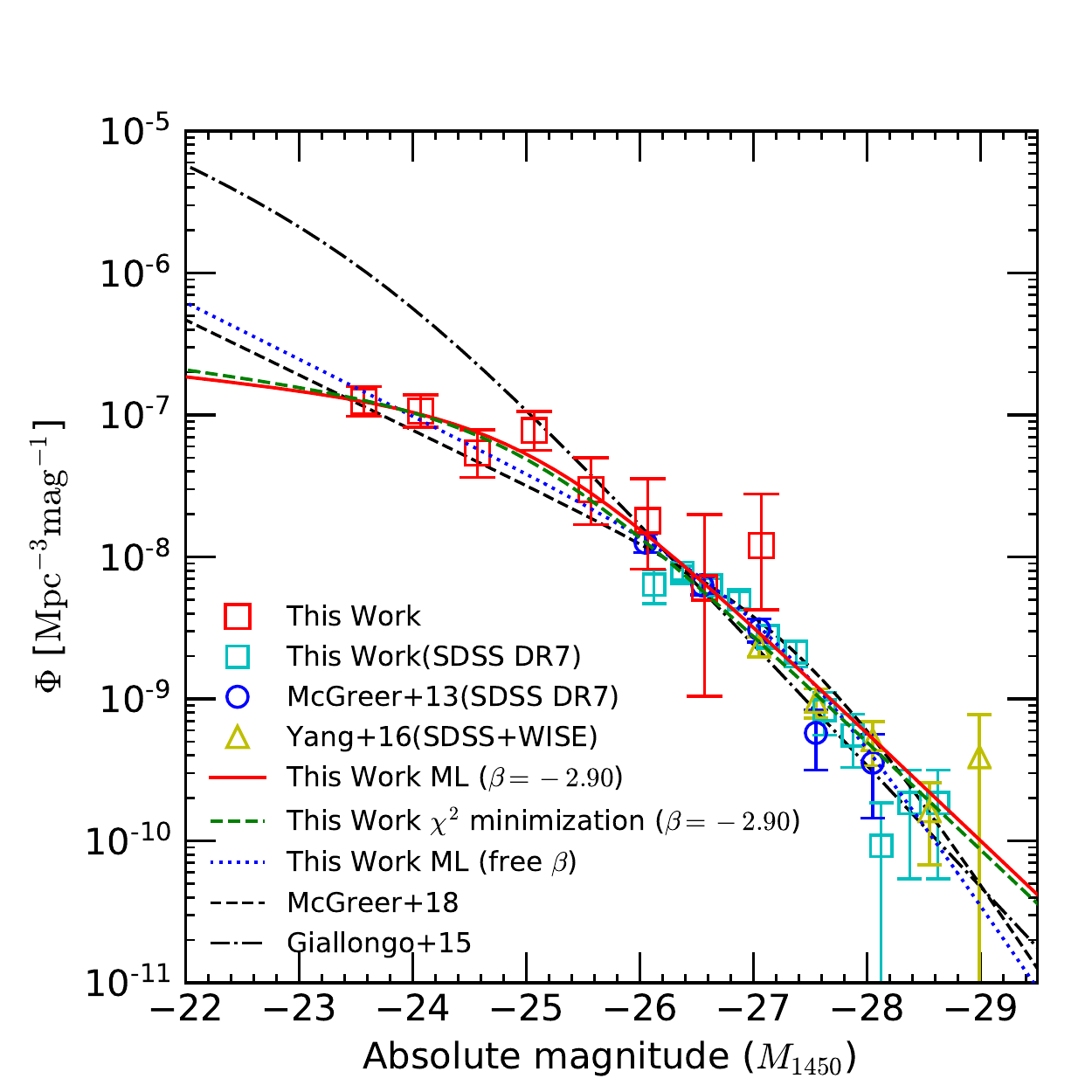}
\caption{
   The QLF at $z \sim 5$. The binned QLF in this work are plotted with red squares (HSC $i < 23.1$) and 
   cyan squares (SDSS DR7). The blue circles show the number densities of \citet{mcg13} from SDSS DR7. 
   The yellow triangles are the quasar number densities at $z = 4.9$, which are converted from $z = 5.1$ 
   by using the redshift evolution given by \citet{fan01}, from SDSS+WISE survey \citep{yan16}.The red 
   solid line and green bashed line show the best-fit double power-law model with the maximum likelihood 
   fit to the our sample and the $\chi^2$ fit to the binned QLF in this work with fixed $\beta$. The blue dotted 
   line show the fitting results of the maximum likelihood fit with free $\beta$. For comparison we also show 
   the best-fit QLF of \citet{mcg18} and \citet{gia15} based on the X-ray detected samples with black 
   bashed line and black point dotted line, respectively.
\label{fig:fig_QLF}}
\end{figure}

\subsection{Double power-law function model} \label{subsec:Double_power-law_model}

The QLF is known to be well described by a double power-law function
\begin{eqnarray}
     \Phi(M_{1450}, z) = \ \ \ \ \ \ \ \ \ \ \ \ \ \ \ \ \ \ \ \ \ \ \ \ \ \ \ \ \ \ \ \ \ \ \ \ \ \ \ \ \ \ \ \ \ \ \ \  \nonumber \\
     \frac{\Phi(M^*_{1450})}
     {10^{0.4(\alpha + 1)(M_{1450} - M^*_{1450}) }+ 10^{0.4(\beta + 1)(M_{1450} - M^*_{1450})}}, 
     \nonumber \\
\end{eqnarray}
where $\alpha$, $\beta$, $\Phi(M^{\ast}_{1450})$, and $M^{\ast}_{1450}$ are the faint-end slope, the 
bright-end slope, the normalization of the QLF, and the characteristic absolute magnitude, respectively 
\citep[e.g.,][]{boy00}. The double power-law model is fitted to the above data, using the maximum 
likelihood method \citep{mar83}. We maximize the likelihood $L$ by minimizing $S = -2 \ln L$, given by
\begin{eqnarray}
   S &=& -2 \sum \ln \Bigl[\Phi_{\rm{p}}(z, M_{1450}) f_{\rm{comp}}(z, M_{1450})\Bigr] 
   \nonumber \\  \nonumber \\
   &+& 2 \int_{-\infty}^{+\infty} \int_{-\infty}^{+\infty} \Phi_{\rm{p}}(z, M_{1450}) f_{\rm{comp}}(z, M_{1450}) 
   dz dM_{1450},\nonumber \\
\end{eqnarray}
where the sum in the first term is taken over all the HSC and SDSS quasars at $M_{1450} < -23.32$. 
Since the spectroscopic redshift of most objects in our HSC sample is unknown, we probabilistically 
assign a redshift to each object in our HSC sample whose redshift distribution should follow the survey 
completeness at the magnitude of each object (Figure~\ref{fig:fig_completeness}). We checked that the 
distribution of assigned redshifts is consistent with the redshift distribution calculated from the best-fit 
luminosity function described bellow. Therefore this assumption seems to be valid. 
We then calculate $M_{1450}$ for each quasar based on the assigned redshift. Consequently, 
for the HSC sample (224 objects with $19.1 < i < 24.1$), the averages with the 
standard deviations of the assigned redshift and $M_{1450}$ are $4.91 \pm 0.25$ and $-23.22 \pm 
1.00$, respectively. For the sub-sample used for the QLF fitting (72 objects in 
$19.1 < i < 23.1$), they are $4.87 \pm 0.22$ and $-24.43 \pm 0.87$. For reference, the average and 
standard deviation of the redshift and $M_{1450}$ for the SDSS spectroscopic sample (288 objects) 
are $4.75 \pm 0.18$ and $-26.78 \pm 0.46$.

The bright-end slope of the QLF at $z \sim 5$ was investigated by \citet{mcg13} and \citet{yan16}. 
\citet{mcg13} used the SDSS DR7 quasar catalog and derived the number densities of $z \sim 5$ 
quasars with optical color selection. \citet{yan16} constructed a sample of bright quasars for a 
magnitude range of $-29.0 < M_{1450} < -26.8$, by combining the SDSS and the Wide-field Infrared 
Survey Explorer (WISE; \citealt{wri10}) data. The sample of \citet{yan16} covers higher luminosities 
than our sample, and contains a larger number of objects than the \citet{mcg13} sample in the brightest 
range. We fit the number densities of $z \sim 5$ bright quasars at $-29.0 < M_{1450} < -25.8$ derived 
by \citet{mcg13} and \citet{yan16} with a single power-law model. The number densities of \citet{yan16} 
at $z = 5.05$ are corrected to $z = 4.9$ by using the redshift evolution given by \citet{fan01}. The 
derived slope is $-2.90^{+0.02}_{-0.03}$. This slope does not change significantly if we limit
the range of the fit to the brighter side ($M_{1450} < -27$) of the data of \citet{mcg13} and \citet{yan16}.
Based on this result we fix the bright-end slope to $\beta = -2.90$ in the following maximum likelihood 
fitting of the QLF. Since the maximum likelihood fitting need the data of redshift and $M_{1450}$ of 
each object and $f_{\rm{comp}}(z, M_{1450})$, we use our SDSS sample described in Section 
\ref{subsec:binnedQLF}, instead of the sample used by \citet{mcg13} and \citet{yan16}.

We summarize the results of our maximum likelihood fitting in the first line of Table \ref{tab:tab_best_para} 
and adopt these values as our best fit. The derived QLF is shown in Figure \ref{fig:fig_QLF}. The fitting 
results show a flat faint-end slope, $\alpha = -1.22^{+0.03}_{-0.10}$. We also fit the binned luminosity 
function with a double power-law model through $\chi^2$ minimization, fixing the bright-end slope to 
$\beta = -2.90$. The best-fit parameters are summarized in the second line of Table \ref{tab:tab_best_para} 
and the fitted QLF is shown in Figure \ref{fig:fig_QLF}. The fitting results in these two cases are consistent 
with each other, and reproduce our HSC number densities reasonably well. We also attempt to fit the QLF 
by the double power-law model by varying $\beta$ as a free parameter, which resulted in a steeper 
faint-end slope ($\alpha = -2.00^{+0.40}_{-0.03}$) as shown in the third line of Table \ref{tab:tab_best_para}. 

\begin{deluxetable*}{ccccc}
\tablecaption{The best-fit parameters of QLF at $z \sim 5$\label{tab:tab_best_para}}
\tablehead{
\colhead{} & \colhead{$\alpha$} & \colhead{$\beta$} & \colhead{$\it{\Phi}(M^{\ast}_{\rm{1450}})$} & \colhead{$M^{\ast}_{\rm{1450}}$} \\
\colhead{} & \colhead{[faint-end]} & \colhead{[bright-end]} & \colhead{(10$^{-7}$ Mpc$^{-3}$ mag$^{-1}$)} & \colhead{(mag)}
}
\startdata
Maximum likelihood (fixed $\beta$) & $-1.22^{+0.03}_{-0.10}$ & $-2.90$ & $1.01^{+0.21}_{-0.29}$ & $-25.05^{+0.10}_{-0.24}$ \\
$\chi^2$ minimization (fixed $\beta$) & $-1.27\pm{0.17}$ & $-2.90$ & $1.00\pm{0.06}$ & $-24.97\pm{0.04}$ \\
Maximum likelihood (free $\beta$) & $-2.00^{+0.40}_{-0.03}$ & $-3.94^{+0.20}_{-0.04}$ & $0.054^{+0.003}_{-0.010}$ & $-27.15^{+0.03}_{-0.10}$ \\
\enddata
\end{deluxetable*}

\section{Discussion} \label{sec:discussion}

\subsection{Comparison with previous measurements} \label{subsec:previous}

\begin{figure*}[ht!]
\includegraphics[width=18.2cm]{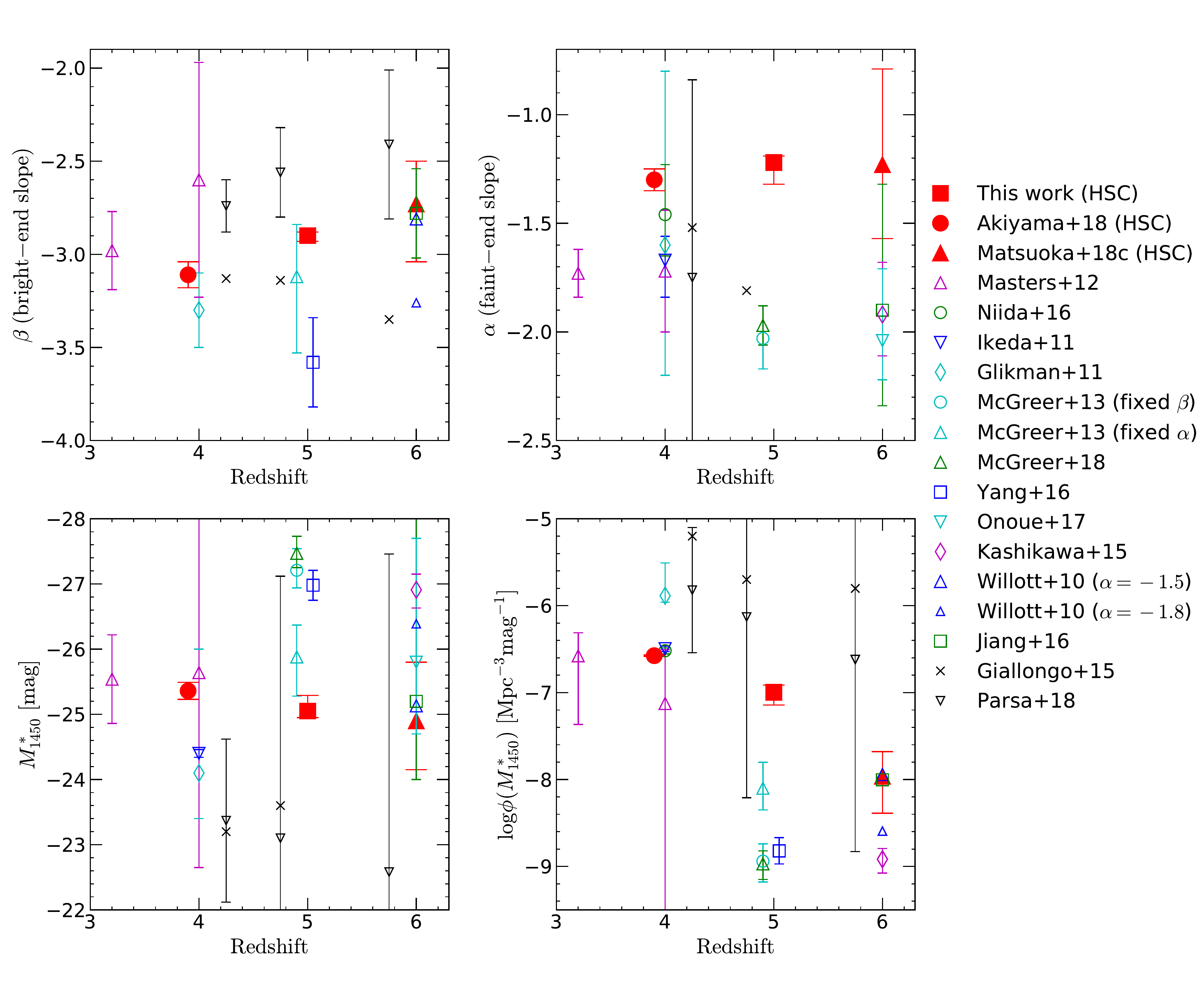}
\caption{
   Best-fit QLF parameters obtained by the present and previous works. Our results by the maximum 
   likelihood method with the fixed $\beta$ are shown by the red filled squares. Red filled circles and 
   triangles: double power-law fit to $z \sim 4$ and $6$ QLF from \citet{aki18} and \citet{mat18c} with 
   HSC survey, respectively. Magenta triangles: double power-law fit to $z \sim 3.2$ and $z \sim 4$ QLF 
   from \citet{mas12}. Green circles and blue reverse triangles: $z \sim 4$ QLF from \citet{nii16} and 
   \citet{ike11}, respectively. Cyan diamonds: $z \sim 4$ QLF from \citet{gli11}. Cyan circles and 
   triangles: fit to $z \sim 5$ QLF from \citet{mcg13} with fixed $\beta$ and $\alpha$, respectively. 
   Green triangles: double power-law fit to $z \sim 5$ QLF from \citet{mcg18} with fixed $\beta$. 
   Blue squares: $z \sim 5$ QLF from \citet{yan16}. Cyan reverse triangles and magenta diamonds: 
   $z \sim 6$ QLF from the case 1 of \citet{ono17} and \citet{kas15}. Blue triangles: bright-end $z \sim 6$ 
   QLF from \citet{wil10} with fixed $\alpha = -1.5$ (large) and $\alpha = -1.8$ (small). Green squares: 
   $z \sim 6$ QLF from \citet{jia16}. Black crosses and reverse triangles: $z = 4.25$, $z = 4.75$, and 
   $z = 5.75$ QLF based on X-ray selected sample from \citet{gia15} and \citet{par18}, respectively. 
\label{fig:fig_LF_para}}
\end{figure*}

We compare our QLF and previous measurements at $z \sim 5$ in Figures \ref{fig:fig_number_density} 
and \ref{fig:fig_QLF}. In Figure \ref{fig:fig_number_density} our binned QLF is compared with the results 
of \citet{mcg18} using the SDSS, the Stripe82 \citep{aba09}, and the CFHTLS \citep{gwy12} data. The 
plotted results are consistent with each other at $M_{1450} < -24.32$. Though our number densities are
systematically higher than those of \citet{mcg18} at the fainter magnitudes, both studies indicate a flat 
faint-end slope. On the other hand, our faint-end slope is flatter than that of \citet{gia15}, $\alpha = -1.81$, 
which is based on X-ray detected samples (see Figures \ref{fig:fig_QLF} and \ref{fig:fig_LF_para}).

We also compare the obtained parameters of the QLF with those of previous studies. This is shown in 
Figure \ref{fig:fig_LF_para}. Our bright-end slope is roughly consistent with those in previous studies, 
because we use essentially the same SDSS sample as used in the previous studies 
\citep{mcg13, mcg18, yan16}. On the other hand, our faint-end slope is flatter than that of the previous 
studies \citep{mcg13, mcg18}. The reason seems to be that we constructed a larger sample of faint 
quasars thanks to the deep and wide HSC-SSP survey data (see also Figure \ref{fig:fig_QLF}). The 
faint-end slope of \citet{mcg18} is steeper than ours, presumably because they fixed the bright-end slope 
to a relatively steep value, $\beta = -4.0$ \citep[see also][]{mat18c}. Our break 
magnitude is fainter and our number density at the break is higher than those of the previous optical 
studies. 

As a result of the comparison with previous optical surveys at other redshifts, it is 
inferred that the bright-end slope is roughly constant, and the number density at the break decreases 
toward higher redshift, in the range of $4 \lesssim z \lesssim 6$. In previous results except for HSC, the 
faint-end slope becomes steeper toward higher redshift. Recent measurements
with HSC-SSP data, e.g., \citet{aki18} and SHELLQs \citep{mat16, mat18a, mat18b, mat18c}, constructed 
large low-luminosity quasar samples and reported the flat faint-end slope of the QLF at $z$ $\sim$ 4 and 6. Once we focus on 
the HSC-SSP results, it is indicated that the faint-end slope and the break magnitude are roughly constant 
for $4 \lesssim z \lesssim 6$. 

\subsection{Evolution of the quasar number density} \label{subsec:evolution}

\begin{figure}[ht!]
\includegraphics[width=8.8cm]{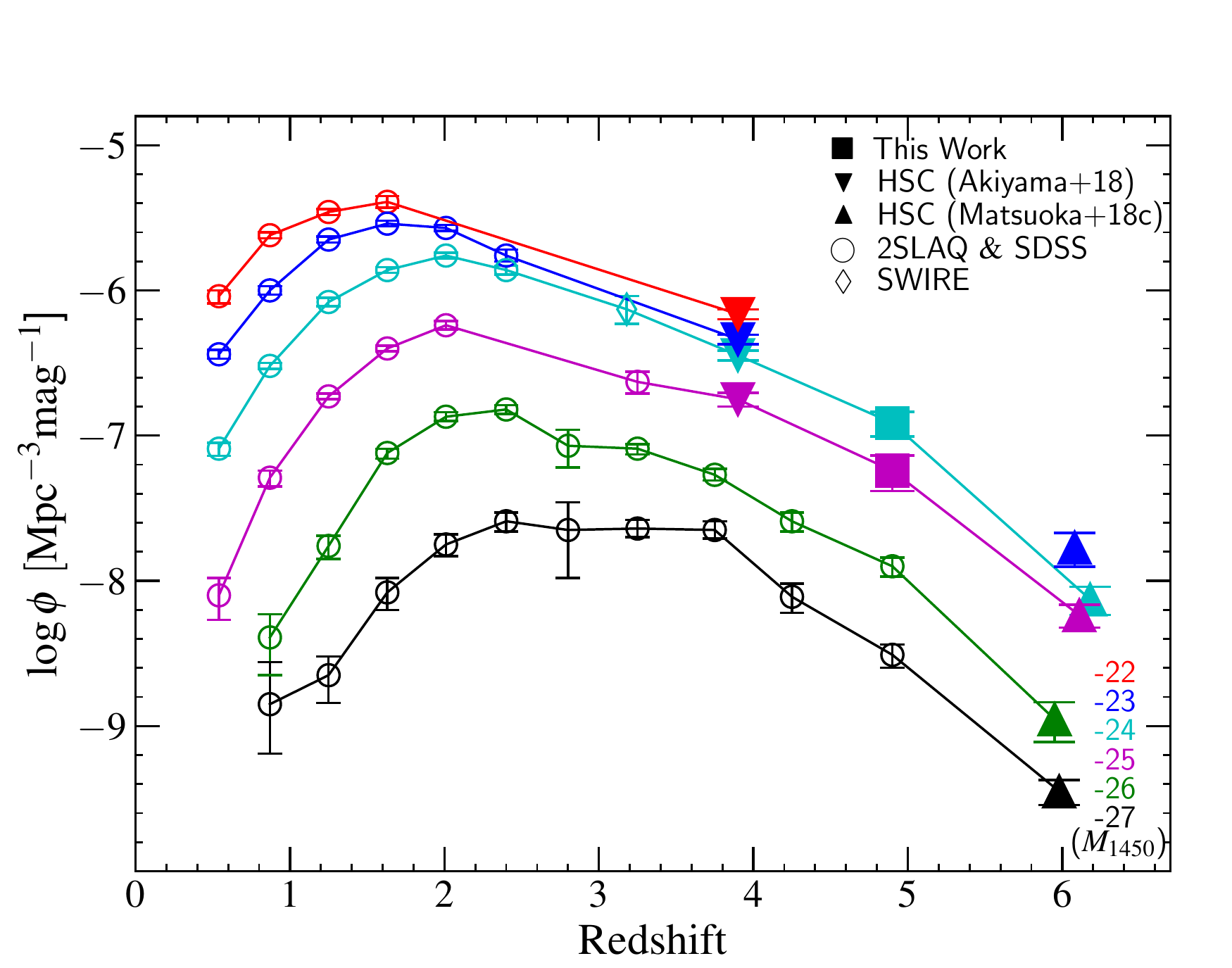}
\caption{
   Redshift evolution of the quasar number density. The red, blue, cyan, magenta, green, and black 
   lines are the number density of quasars with $M_{1450} = -22$, $-23$, $-24$, $-25$, $-26$, and 
   $-27$, respectively. The filled squares, reverse triangles, and triangles show the HSC results of this 
   study, \citet{aki18}, and \citet{mat18c}, respectively. The open circles and the open diamond 
   represent the results obtained in the 2dFSDSS LRG and Quasar Survey \citep[2SLAQ;][]{cro09} 
   and SDSS \citep{ric06, mcg13}, and the Spitzer Wide-area Infrared Extragalactic Legacy Survey 
   \citep[SWIRE;][]{sia08}, respectively.
\label{fig:fig_evolution_phi}}
\end{figure}

The number density of low-luminosity quasars at high redshifts is a key to understanding 
luminosity-dependent evolution of quasars and to discuss the cosmological evolution of SMBHs at 
different masses. Figure \ref{fig:fig_evolution_phi} shows the quasar number density 
at different absolute magnitudes as a function of redshift. Based on the quasar luminosity function at 
$z \sim 5$ derived in Section \ref{subsec:binnedQLF}, we calculate the quasar 
number densities at $M_{1450} = -24$ and $-25$. These magnitude ranges are not 
affected by the contamination as described in Section \ref{subsec:contamination} and 
\ref{subsec:Double_power-law_model}. At $z > 3$, previous optical studies indicate 
that the number density of low-luminosity quasars increases toward higher redshift \citep{gli10, gli11}. 
On the contrary, the HSC results presented in this paper, \citet{aki18}, and \citet{mat18c} suggest that 
the number densities of faint quasars decrease toward high redshift. The reason for the discrepancy 
between the \citet{gli10, gli11} and present work is unclear, but may be at least in part due to different 
selection criteria, e.g., the point source separation.

We further explore the density evolution of quasars in the range of $4 \leq z \leq 6$. 
Some recent studies \citep{jia16, wan18, yan18} reported a more rapid decline in the number density of 
luminous quasars toward higher redshift at $z \geq 4$. We fit an exponentially-declining 
function to the number density of low-luminosity quasars ($M_{1450} = -24$ and $-25$) evaluated from 
the HSC survey data,
\begin{eqnarray}
   \Phi(z, M_{1450}) =  \Phi(z = 4.9, M_{1450}) 10^{k(z - 4.9)}. \label{eqa:k}
\end{eqnarray} 
The derived density evolution parameter ($k$ in Equation \ref{eqa:k}) for quasars with 
$M_{1450} = -24$ is $k = -0.47$ at $z \sim 4-5$ and $k = -0.95$ at $z \sim 5-6$. The 
parameter for quasars with $M_{1450} = -25$ is $k = -0.49$ at $z \sim 4-5$ and 
$k = -0.82$ at$z \sim 5-6$. Therefore, as seen in luminous quasars \citep{jia16, wan18, yan18}, we 
confirmed that the number density of low-luminosity quasars decreases more rapidly 
from $z \sim 5$ to $z \sim 6$ than from $z \sim 4$ to $z \sim 5$. Though the $k$ 
parameter shows a clear redshift dependence, it does not show a significant luminosity dependence at 
a fixed redshift range (Figure~\ref{fig:fig_evolution_phi}).

\section{Summary} \label{sec:summary}

This paper presented the QLF at $z \sim 5$ in a wide luminosity range. We constructed a statistical 
sample of 224 $z \sim 5$ quasars in the magnitude range of $19.1 < i < 24.1$, based on the 
$g$, $r$, $i$, and $y$-band imaging data over 81.8 deg$^2$ taken from the 
S16A-Wide2 release of the HSC-SSP survey. The quasar candidates were selected by their point-source 
morphology and $r$-band dropout feature. With estimates of survey completeness and effective area, we 
calculated the binned QLF. We fitted a double power-law function model to the sample 
at $M_{1450} < -23.32$ where the effect of contamination is minimal. The main results of this study are 
summarized as follows.
 
\begin{enumerate}
    \item \ \ Our binned QLF at $z \sim 5$ is consistent with those of previous studies at $M_{1450} < -24.32$ 
    \citep[e.g.,][]{mcg13, mcg18, yan16}. 
    \item \ \ We got the best-fit faint-end slope $\alpha = -1.22^{+0.03}_{-0.10}$ when the bright-end slope 
    was fixed to $\beta = -2.90$. This is flatter than those reported in previous studies at $z \sim 5$. The 
    break of the double power-law is fainter and the number densities at the break is higher than reported in 
    the previous studies. 
     \item \ \ Combined with the HSC results at $z = 4$ and $z = 6$, our results suggest that there is little 
     redshift evolution of the break magnitude and the faint-end slope at $4 \leq z \leq 6$. On the other hand, 
     the number density at the break decreases toward high redshift. 	
    \item \ \ The number density of low-luminosity quasars decreases toward high redshift at $z > 3$. The 
    number density of low-luminosity quasars decreases more rapidly from $z \sim 5$ to $z \sim 6$ than from 
    $z \sim 4$ to $z \sim 5$ as also seen in luminous quasars \citep{jia16, wan18, yan18}. 
\end{enumerate}

\acknowledgments

We thank the anonymous referee for valuable comments, which improved this paper significantly.

The Hyper Suprime-Cam (HSC) collaboration includes the astronomical communities of Japan and Taiwan, and Princeton University.  The HSC instrumentation and software were developed by the National Astronomical Observatory of Japan (NAOJ), the Kavli Institute for the Physics and Mathematics of the Universe (Kavli IPMU), the University of Tokyo, the High Energy Accelerator Research Organization (KEK), the Academia Sinica Institute for Astronomy and Astrophysics in Taiwan (ASIAA), and Princeton University.  Funding was contributed by the FIRST program from Japanese Cabinet Office, the Ministry of Education, Culture, Sports, Science and Technology (MEXT), the Japan Society for the Promotion of Science (JSPS),  Japan Science and Technology Agency  (JST),  the Toray Science  Foundation, NAOJ, Kavli IPMU, KEK, ASIAA,  and Princeton University.

The Pan-STARRS1 Surveys (PS1) have been made possible through contributions of the Institute for Astronomy, the University of Hawaii, the Pan-STARRS Project Office, the Max-Planck Society and its participating institutes, the Max Planck Institute for Astronomy, Heidelberg and the Max Planck Institute for Extraterrestrial Physics, Garching, The Johns Hopkins University, Durham University, the University of Edinburgh, Queen's University Belfast, the Harvard-Smithsonian Center for Astrophysics, the Las Cumbres Observatory Global Telescope Network Incorporated, the National Central University of Taiwan, the Space Telescope Science Institute, the National Aeronautics and Space Administration under Grant No. NNX08AR22G issued through the Planetary Science Division of the NASA Science Mission Directorate, the National Science Foundation under Grant No. AST-1238877, the University of Maryland, and Eotvos Lorand University (ELTE).

This paper makes use of software developed for the Large Synoptic Survey Telescope. We thank the LSST Project for making their code available as free software at http://dm.lsst.org.

This work is based on data collected at the Subaru Telescope and retrieved from the HSC data archive system, which is operated by the Subaru Telescope and Astronomy Data Center at National Astronomical Observatory of Japan.

Data analysis was in part carried out on the Multi-wavelength Data Analysis System operated by the Astronomy Data Center (ADC), National Astronomical Observatory of Japan.

Funding for the Sloan Digital Sky Survey IV has been provided by the Alfred P. Sloan Foundation, the U.S. Department of Energy Office of Science, and the Participating Institutions. SDSS-IV acknowledges
support and resources from the Center for High-Performance Computing at
the University of Utah. The SDSS web site is www.sdss.org.

SDSS-IV is managed by the Astrophysical Research Consortium for the 
Participating Institutions of the SDSS Collaboration including the 
Brazilian Participation Group, the Carnegie Institution for Science, 
Carnegie Mellon University, the Chilean Participation Group, the French Participation Group, Harvard-Smithsonian Center for Astrophysics, 
Instituto de Astrof\'isica de Canarias, The Johns Hopkins University, 
Kavli Institute for the Physics and Mathematics of the Universe (IPMU) / 
University of Tokyo, the Korean Participation Group, Lawrence Berkeley National Laboratory, 
Leibniz Institut f\"ur Astrophysik Potsdam (AIP),  
Max-Planck-Institut f\"ur Astronomie (MPIA Heidelberg), 
Max-Planck-Institut f\"ur Astrophysik (MPA Garching), 
Max-Planck-Institut f\"ur Extraterrestrische Physik (MPE), 
National Astronomical Observatories of China, New Mexico State University, 
New York University, University of Notre Dame, 
Observat\'ario Nacional / MCTI, The Ohio State University, 
Pennsylvania State University, Shanghai Astronomical Observatory, 
United Kingdom Participation Group,
Universidad Nacional Aut\'onoma de M\'exico, University of Arizona, 
University of Colorado Boulder, University of Oxford, University of Portsmouth, 
University of Utah, University of Virginia, University of Washington, University of Wisconsin, 
Vanderbilt University, and Yale University.

This work is based in part on data collected at Subaru Telescope, which is operated by the National Astronomical Observatory of Japan.
We appreciate the staff members of the telescope for their support during our FOCAS observations.

Based on observations at Cerro Tololo Inter-American Observatory, National Optical Astronomy Observatory (NOAO Prop. ID 2016A-0395; PI: M. Niida), which is operated by the Association of Universities for Research in Astronomy (AURA) under a cooperative agreement with the National Science Foundation.
We appreciate the staff members of the telescope for their support during our CTIO observations.

This work was financially supported in part by JSPS (MN: 18J11887; TN: 16H03958, 17H01114, 19H00697, and 20H01949; YT: 18J01050; YM: 17H04830).
MN acknowledges support also from the Hayakawa Satio Fund of Astronomical Society of Japan. 
TN was supported also by the grant of the NAOJ Visiting Joint Research program of the NAOJ Research Coordination Committee.
YM was supported also by the Mitsubishi Foundation Grant No. 30140.

\appendix

\section{Spectroscopic observations} \label{sec:spec}

We conducted spectroscopic observations toward six candidates of $z \sim 5$ quasars as a pilot study 
for more systematic programs in the future. The targets were selected 
based on the visibility at the time of observations and their apparent magnitudes. The 
observations were conducted by the 4m Blanco telescope in the Cerro Tololo Inter-American Observatory 
(CTIO) and the 8.2m Subaru Telescope of the National Astronomical Observatory of Japan. Among the six 
objects, four objects meet our latest quasar selection criteria described in this paper, while the remaining 
two objects do not (see Table \ref{tab:tab_target}). 

\subsection{Subaru Telescope} \label{subsec:Subaru Telescope}

We observed two faint ($i > 23$) objects with the Subaru Telescope, in which one object (J0205--0353) 
meets our latest quasar selection criteria. The observations were carried out with the Faint Object Camera 
and Spectrograph \citep[FOCAS;][]{kas02} on 2016 October 7 (S16B-0180S; Niida et al.). We used the 
300R grating with the SO58 filter whose wavelength coverage is 
$5800 < \lambda ($\AA$) < 10000$. A $0.\!\!\arcsec8$ width longslit was used, resulting in a wavelength 
resolution of $R$ $\sim$ 900. The seeing size was $\sim$$0.\!\!\arcsec7$ on average. The individual 
exposure time was $650 - 1050$ s, and the total exposure time was $2700 - 2750$ s for each object.

\begin{figure*}[ht!]
\includegraphics[width=17.5cm]{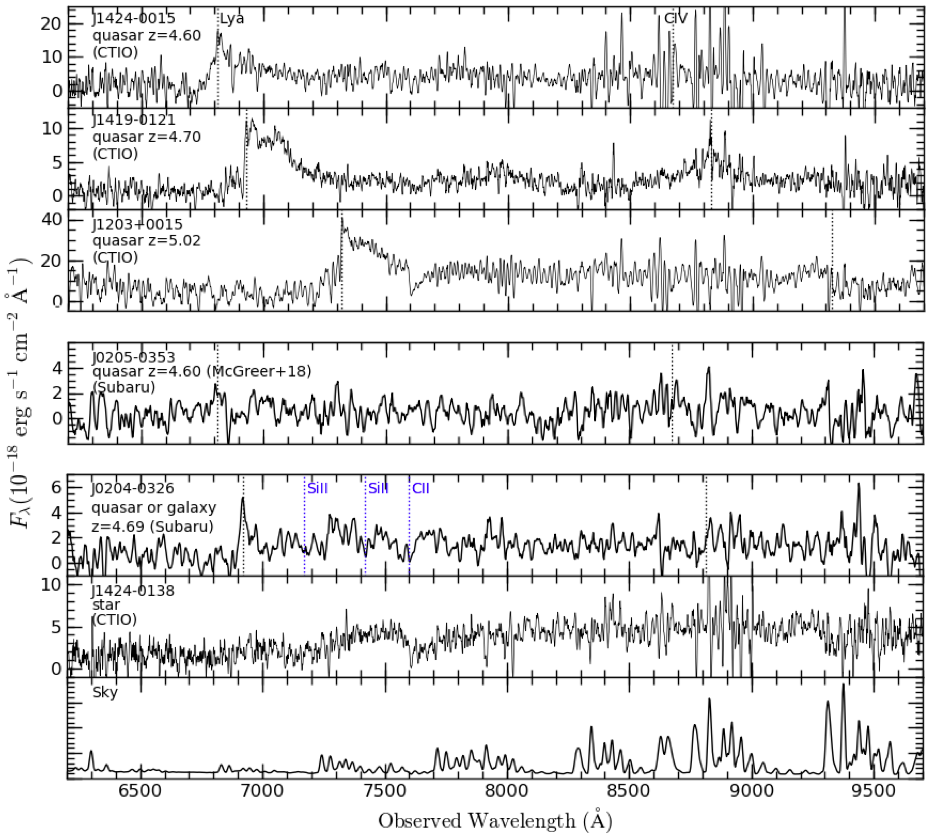}
\caption{
   Reduced spectra of the 6 objects, obtained with Subaru/FOCAS and CTIO/COSMOS. The spectra of
   the objects, which are confirmed as quasars by our observations, are shown at the 
   top three panels. Fourth panel shows the spectrum of a quasar, which was confirmed 
   by \citet{mcg18}. Fifth and sixth panels show the spectra of objects that does not satisfy our quasar 
   selection criteria. The object name, the measured redshift and the telescope used to take the spectrum 
   are indicated at the top left corner of the each panel. The black dotted lines represent the expected 
   positions of the Ly$\alpha$ $\lambda1216$ and C~{\sc iv} $\lambda1549$ emission lines. The expected 
   positions of the interstellar absorption lines of Si~{\sc ii} $\lambda$1260, Si ~{\sc ii} $\lambda$1304, and 
   C~{\sc ii} $\lambda$1335 are marked by the blue dotted lines for J0204--0326. The spectra are smoothed 
   with $5 - 7$ pixel boxcar. The bottom panel displays a typical sky spectrum with an arbitrary flux scale.
\label{fig:fig_spec}}
\end{figure*}

We used IRAF for the data reduction. The flux calibration was tied to a spectrophotometric standard star, 
Feige 110. The targets are listed in Table \ref{tab:tab_target} and their reduced spectra are shown in 
Figure \ref{fig:fig_spec}. J0205--0353 shows only a featureless continuum, while this object was identified 
as a $z = 4.60$ quasar by \citet{mcg18} in a better seeing condition. J0204--0326 shows a relatively 
narrow Ly$\alpha$ emission line, $v_{\rm{FWHM}}(\rm{Ly}\alpha) \sim 1010$ km s$^{-1}$. This object 
dose not satisfy the point source selection criteria defined by Equation (\ref{equ2}) nor $g$ and $r$ 
{\tt countinputs} $\geq 4$, though it satisfies the remaining selection criteria. Its spectrum 
shows weak features of the interstellar absorption lines of Si~{\sc ii} $\lambda$1260, 
Si ~{\sc ii} $\lambda$1304, and C~{\sc ii} $\lambda$1335, which are shown in Figure \ref{fig:fig_spec}. 
Therefore we classified this object as a $z = 4.69$ quasar or galaxy.

\begin{deluxetable}{lcccccccc}
\tablecaption{The list of objects for the spectroscopic observations\label{tab:tab_target}}
\tablehead{
    \colhead{Name} & 
    \colhead{$r_{\rm{AB}}$} & 
    \colhead{$i_{\rm{AB}}$ } & 
    \colhead{$y_{\rm{AB}}$} & 
    \colhead{Type} & 
    \colhead{Redshift} & 
    \colhead{$M_{1450}$} & 
    \colhead{Exp. Time (s)} & 
    \colhead{Telescope} 
}
\startdata
    HSC J020435.29--032654.4\tablenotemark{a} & $25.13$ & 23.66 & $23.51$ & 
        quasar or galaxy & 4.69 & --22.67 & 2700 & Subaru\\
    HSC J020541.60--035350.8 & $24.25$ & 23.15 & $23.08$ & 
        quasar & 4.60\tablenotemark{b} & --23.14 & 2750 & Subaru\\
    HSC J120343.47+001527.4 & $22.23$ & 20.61 & $20.33$ & 
        quasar & 5.02 & --25.86 & 9900 & CTIO\\
    HSC J141943.69--012114.6 & $22.75$ & 21.61 & $21.35$ & 
        quasar & 4.70 & --24.72 & 6300 & CTIO\\
    HSC J142437.92--001503.0 & $22.60$ & 21.35 & $21.02$ & 
        quasar & 4.60 & --24.94 & 7200 & CTIO\\
    HSC J142421.20--013827.3\tablenotemark{a} & $22.75$ & 21.26 & $21.35$ & 
        star & $\cdots$ & $\cdots$ & 2700 & CTIO\\
\enddata
\tablenotetext{a}{
    These objects are not selected as $z \sim 5$ quasar candidates by our latest criteria presented in this work.}
\tablenotetext{b}{
    This redshift is measured by \citet{mcg18}.}
\end{deluxetable}

\subsection{CTIO} \label{subsec:CTIO}

We observed four bright objects, $i < 23$, with the 4m Blanco telescope at CTIO. Three of them meet our 
latest quasar selection criteria, while the remaining one does not. The observations were 
carried out with the Cerro Tololo Ohio State Multi-Object Spectrograph (COSMOS) on 2016 April 12--15 
(2016A-0395; Niida et al.). We used the Red VPH Grism with the OG570 filter whose wavelength coverage 
is $6110 < \lambda ($\AA$) < 10275$. A $1.\!\!\arcsec2$ width longslit was used, resulting in a wavelength 
resolution of $R$ $\sim$ 1900. The seeing size was $\sim$$1.\!\!\arcsec1- 1.\!\!\arcsec3$ on average. 
The individual exposure time was $900$ s, and the total exposure time was $2700 - 9900$ s for each object.

We used IRAF for the data reduction. The flux calibration was tied to spectrophotometric standard stars, 
Hiltner 600 and LTT 6248. The targets are listed in Table \ref{tab:tab_target} and their reduced spectra are 
shown in Figure \ref{fig:fig_spec}. All of the three candidates meeting our latest quasar selection criteria 
were identified as $z \sim 5$ quasars. The remaining object (J1424--0138) satisfies all the criteria but their 
$g$ and $r$ countinputs $< 4$. It has red continuum and dose not show a Lyman break or broad emission 
lines. Therefore we conclude that it is a Galactic star.


\begin{thebibliography}{}
\bibitem[Abazajian et al.(2009)]{aba09} Abazajian, K. N., Adelman-McCarthy, J. K., Ag\"{u}eros, M. A., et al. 2009, \apjs, 182, 543
\bibitem[Adams et al.(2019)]{ada19} Adams, N. J., Bowler, R. A. A., Jarvis, M. J., et al. 2019, arXiv:1912.01626
\bibitem[Aihara et al.(2018a)]{aih18a} Aihara, H., Arimoto, N., Armstrong, R., et al. 2018a, \pasj, 70, S4
\bibitem[Aihara et al.(2018b)]{aih18b} Aihara, H., Armstrong, R., Bickerton, S., et al. 2018b, \pasj, 70, S8
\bibitem[Aird et al.(2015)]{air15} Aird, J., Coil, A. L., Georgakakis, A., et al. 2015, \mnras, 451, 1892
\bibitem[Akiyama et al.(2018)]{aki18} Akiyama, M., He, W., Ikeda, H., et al. 2018, \pasj, 70, S34
\bibitem[Baldwin(1977)]{bal77} Baldwin, J. A. 1977, \apj, 214, 679
\bibitem[Ba\~nados et al.(2018)]{ban18} Ba\~nados, E., Venemans, B. P., Mazzucchelli, c., et al. 2018, Nature, 553, 473
\bibitem[Baskin \& Laor(2004)]{bas04} Baskin, A., \& Laor, A. 2004, ASPC, 311, 175
\bibitem[Bosch et al.(2018)]{bos18} Bosch, J., Armstrong, R., Bickerton, S., et al. 2018, \pasj, 70, S5
\bibitem[Boutsia et al.(2018)]{bou18} Boutsia, K., Grazian, A., Giallongo, E., Fiore, F., \& Civano, F. 2018, \apj, 869, 20
\bibitem[Boyle et al.(2000)]{boy00} Boyle, B. J., Shanks, T., Croom, S. M., et al. 2000, \mnras, 317, 1014
\bibitem[Chabrier et al.(2003)]{cha03} Chabrier, G. 2003, \pasp, 115, 763
\bibitem[Chambers et al.(2016)]{cha16} Chambers, K. C., Magnier, E. A., Metcalfe, N., et al. 2016, arXiv:1612.05560
\bibitem[Coupon et al.(2018)]{cou18} Coupon, J., Czakon, N., Bosch, J., et al. 2018, \pasj, 70, S7
\bibitem[Cowie et al.(1996)]{cow96} Cowie, L. L., Songaila, A., Hu, E. M., \& Cohen, J. G. 1996, \aj, 112, 839
\bibitem[Croom et al.(2009)]{cro09} Croom, S. M., Richards, G. T., Shanks, T., et al. 2009, \mnras, 399, 1755
\bibitem[Enoki et al.(2014)]{eno14} Enoki, M., Ishiyama, T., Kobayashi, M. A. R., \& Nagashima, M. 2014, \apj, 794, 69 
\bibitem[Fan et al.(2001)]{fan01} Fan, X., Strauss, M. A., Schneider, D. P., et al. 2001, \aj, 121, 54
\bibitem[Fontanot et al.(2009)]{fon09} Fontanot, F., De Lucia, G., Monaco, P., Somerville, R. S., \& Santini, P. 2009, \mnras, 397, 1776
\bibitem[Francis(1996)]{fra96} Francis, P. J. 1996, \pasa, 13, 212
\bibitem[Furusawa et al.(2018)]{fur18} Furusawa, H., Koike, M., Takata, T., et al. 2018, \pasj, 70, S3
\bibitem[Gehrels(1986)]{geh86} Gehrels, N. 1986, \apj, 303, 336
\bibitem[Giallongo et al.(2015)]{gia15} Giallongo, E., Grazian, A., Fiore, F., et al. 2015, \aap, 578, A83
\bibitem[Girardi et al.(2005)]{gir05} Girardi, L., Groenewegen, M. A. T., Hatziminaoglou, E., \& da Costa, L. 2005, \aap, 436, 895
\bibitem[Glikman et al.(2010)]{gli10} Glikman, E., Bogosavljevi\'c, M., Djorgovski, S. G., et al. 2010, \apj, 710, 1498
\bibitem[Glikman et al.(2011)]{gli11} Glikman, E., Djorgovski, S. G., Stern, D., et al. 2011, \apj, 728, L26
\bibitem[G\"ultekin et al.(2009)]{gul09} G\"ultekin, K., Richstone, D. O., Gebhardt, K., et al. 2009, \apj, 698, 198
\bibitem[Gunn \& Stryker(1983)]{gun83} Gunn, J. E., \& Stryker, L. L. 1983, \apjs, 52, 121
\bibitem[Gwyn(2012)]{gwy12} Gwyn, S. D. J. 2012, \aj, 143, 38
\bibitem[H\"aring \& Rix(2004)]{hari04} H\"aring, N., \& Rix, H.-W. 2004, \apj, 604, L89
\bibitem[Hasinger et al.(2005)]{has05} Hasinger, G., Miyaji, T., \& Schmidt, M. 2005, \aap, 441, 417
\bibitem[Hirata \& Seljak(2003)]{hir03} Hirata, C., \& Seljak, U. 2003, \mnras, 343, 459
\bibitem[Ikeda et al.(2011)]{ike11} Ikeda, H., Nagao, T., Matsuoka, K., et al. 2011, \apj, 728, L25
\bibitem[Ikeda et al.(2012)]{ike12} Ikeda, H., Nagao, T., Matsuoka, K., et al. 2012, \apj, 756, 160
\bibitem[Inoue \& Iwata(2008)]{ino08} Inoue, A. K., Iwata, I., 2008, \mnras, 387, 1681
\bibitem[Inoue et al.(2014)]{ino14} Inoue, A. K., Shimizu, I., Iwata, I., \& Tanaka, M. 2014, \mnras, 442, 1805
\bibitem[Jiang et al.(2006)]{jia06} Jiang, L., Fan, X., Hines, D. C., et al. 2006, \aj, 132, 2127
\bibitem[Jiang et al.(2016)]{jia16} Jiang, L., McGreer, I. D., Fan, X., et al. 2016, \apj, 833, 222
\bibitem[Kashikawa et al.(2002)]{kas02} Kashikawa, N., Aoki, K., Asai, R., et al. 2002, \apj, 54, 819
\bibitem[Kashikawa et al.(2015)]{kas15} Kashikawa, N., Ishizaki, Y., Willott, C. J., et al. 2015, \apj, 798, 28
\bibitem[Kawanomoto et al.(2018)]{kaw18} Kawanomoto, S., Uraguchi, F., Komiyama, Y., et al. 2018, \pasj, 70, 66
\bibitem[Kawara et al.(1996)]{kaw96} Kawara, K., Murayama, T., Taniguchi, Y., \& Arimoto, N. 1996, \apj, 470, L85
\bibitem[Kinney et al.(1990)]{kin90} Kinney, A. L., Rivolo, A. R., \& Koratkar, A. R. 1990, \apj, 357, 338
\bibitem[Komiyama et al.(2018)]{kom18} Komiyama, Y., Obuchi, Y., Nakaya, H., et al. 2018, \pasj, 70, S2
\bibitem[Kuhn et al.(2001)]{kuh01} Kuhn, O., Elvis, M., Bechtold, J., \& Elston, R. 2001, \apjs, 136, 225
\bibitem[Kurk et al.(2007)]{kur07} Kurk, J. D., Walter, F., Fan, X., et al. 2007, \apj, 669, 32
\bibitem[Lusso et al.(2012)]{lus12} Lusso, E., Comastri, A., Simmons, B. D., et al. 2012, \mnras, 425, 623
\bibitem[Marconi \& Hunt(2003)]{mahu03} Marconi, A., \& Hunt, L. K. 2003, \apj, 589, L21
\bibitem[Marshall et al.(1983)]{mar83} Marshall, H. L., Tananbaum, H., Avni, Y., \& Zamorani, G. 1983, \apj, 269, 35
\bibitem[Masters et al.(2012)]{mas12} Masters, D., Capak, P., Salvato, M., et al. 2012, \apj, 755, 169
\bibitem[Matsuoka et al.(2011)]{mat11} Matsuoka, K., Nagao, T., Marconi, A., Maiolino, R., \& Taniguchi, Y. 2011, \aap, 527, A100
\bibitem[Matsuoka et al.(2016)]{mat16} Matsuoka, Y., Onoue, M., Kashikawa, N., et al. 2016, \apj, 828, 26
\bibitem[Matsuoka et al.(2018a)]{mat18a} Matsuoka, Y., Onoue, M., Kashikawa, N., et al. 2018a, \pasj, 70, S35
\bibitem[Matsuoka et al.(2018b)]{mat18b} Matsuoka, Y., Iwasawa, K., Onoue, M., et al. 2018b, \apjs, 237, 5
\bibitem[Matsuoka et al.(2018c)]{mat18c} Matsuoka, Y., Strauss, M. A., Kashikawa, N., et al. 2018c, \apj, 869, 150
\bibitem[Matute et al.(2013)]{mat13} Matute, I., Masegosa, J., M\'{a}rquez, I., et al. 2013, \aap, 557, A78
\bibitem[McGreer et al.(2013)]{mcg13} McGreer, I. D., Jiang, L., Fan, X., et al. 2013, \apj, 768, 105
\bibitem[McGreer et al.(2018)]{mcg18} McGreer, I. D., Fan, X., Jiang, L., \& Cai, Z. 2018, \aj, 155, 131
\bibitem[Miyaji et al.(2015)]{miy15} Miyaji, T., Hasinger, G., Salvato, M., et al. 2015, \apj, 804, 104
\bibitem[Miyazaki et al.(2018)]{miy18} Miyazaki, S., Komiyama, Y., Kawanomoto, S., et al. 2018, \pasj, 70, S1
\bibitem[Mortlock et al.(2011)]{mor11} Mortlock, D. J., Warren, S. J., Venemans, B. P., et al. 2011, Natur, 474, 616
\bibitem[Nagao et al.(2006)]{nag06}  Nagao, T., Marconi, A., \& Maiolino, R. 2006, \aap, 447, 157
\bibitem[Neistein et al.(2006)]{nei06} Neistein, E., van den Bosch, F. C., \& Dekel, A. 2006, \mnras, 372, 933
\bibitem[Niida et al.(2016)]{nii16} Niida, M., Nagao, T., Ikeda, H., et al. 2016, \apj, 832, 208
\bibitem[Ono et al.(2018)]{ono18} Ono, Y., Ouchi, M., Harikane, Y., et al. 2018, \pasj, 70, S10
\bibitem[Onoue et al.(2019)]{ono19} Onoue, M., Kashikawa, N., Matsuoka, Y., et al. 2019, \apj, 880, 77
\bibitem[Onoue et al.(2017)]{ono17} Onoue, M., Kashikawa, N., Willott, C. J., et al. 2017, \apj, 847, L15
\bibitem[Palanque-Delabrouille et al.(2013)]{pal13} Palanque-Delabrouille, N., Magneville, Ch., Y\`eche, Ch., et al. 2013, \aap, 551, A29
\bibitem[P\^aris et al.(2017)]{par17} P\^aris, I., Petitjean, P., Ross, N.P., et al. 2017, \aap, 597, 79
\bibitem[Parsa et al.(2018)]{par18} Parsa, S., Dunlop, J.S., \& McLure, R.J. 2018, \mnras, 474, 2904
\bibitem[Rees(1984)]{ree84} Rees, M. J. 1984, \araa , 22, 471
\bibitem[Pickles(1998)]{pic98} Pickles, A. J. 1998, \pasp, 110, 863
\bibitem[Richards et al.(2006)]{ric06} Richards, G. T., Strauss, M. A., Fan, X., et al. 2006, \aj, 131, 2766
\bibitem[Ross et al.(2013)]{ros13} Ross, N. P., McGreer, I. D., White, M., et al. 2013, \apj, 773, 14
\bibitem[Schlegel et al.(1998)]{sch98} Schlegel, D. J., Finkbeiner, D. P., Davis, M., et al. 1998, \apj, 500, 525
\bibitem[Schneider et al.(2010)]{sch10} Schneider, D. P., Richards, G. T., Hall, P. B., et al. 2010, \aj, 139, 2360
\bibitem[Serjeant et al.(2000)]{ser00} Serjeant, S., Oliver, S., Rowan-Robinson, M., et al. 2000, \mnras, 316, 768
\bibitem[Shen et al.(2008)]{she08} Shen, Y., Greene, J. E., Strauss, M. A., et al. 2008, \apj, 680, 169
\bibitem[Shen et al.(2011)]{she11} Shen, Y., Richards, G. T., Strauss, M. A., et al. 2011, \apjs, 194, 45
\bibitem[Shen \& Kelly(2012)]{she12} Shen, Y., \& Kelly, B. C. 2012, \apj, 746, 169
\bibitem[Shen et al.(2019)]{she19} Shen, Y., Wu, J., Jiang, L., et al. 2019, \apj, 873, 35
\bibitem[Shirakata et al.(2019)]{shi19} Shirakata, H., Okamoto, T., Kawaguchi, T., et al. 2019, \mnras, 482, 4846
\bibitem[Siana et al.(2008)]{sia08} Siana, B., Polletta, M. d. C., Smith, H. E., et al. 2008, \apj, 675, 49
\bibitem[Stevans et al.(2018)]{ste18} Stevans, M. L., Fenkelstein, S. L., Wold, I., et al. 2018, \apj, 863, 63
\bibitem[Takada et al.(2014)]{tak14} Takada, M., Ellis, R. S., Chiba, M., et al. 2014, \pasj, 66, R1
\bibitem[Telfer et al.(2002)]{tel02} Telfer, R. C., Zheng, W., Kriss, G. A., \& Davidsen, A. F. 2002, \apj, 565, 773
\bibitem[Trump et al.(2011)]{tru11} Trump, J. R., Impey, C. D., Kelly, B. C., et al. 2011, \apj, 733, 60
\bibitem[Ueda et al.(2003)]{ued03} Ueda, Y., Akiyama, M., Ohta, K., \& Miyaji, T. 2003, \apj, 598, 886
\bibitem[Ueda et al.(2014)]{ued14} Ueda, Y., Akiyama, M., Hasinger, G., Miyaji, T., \& Watson, M. G. 2014, \apj, 786, 104
\bibitem[Vanden Berk et al.(2001)]{van01} Vanden Berk, D. E., Richards, G. T., Bauer, A., et al. 2001, \aj, 122, 549
\bibitem[Venemans et al.(2013)]{ven13} Venemans, B. P., Findlay, J. R., Sutherland, W. J., et al. 2013, \apj, 779, 24 
\bibitem[Venemans et al.(2015)]{ven15} Venemans, B. P., Ba\~{n}ados, E., Decarli, R., et al. 2015, \apj, 801, L11
\bibitem[Vestergaard \&Peterson(2006)]{ves06} Vestergaard, M., \& Peterson, B. M. 2006, \apj, 641, 689
\bibitem[Wang et al.(2019)]{wan18} Wang, F., Yang, J., Fan, X., et al. 2019, \apj, 884, 30
\bibitem[Willott et al.(2010)]{wil10} Willott, C. J., Delorme, P., Reyl\'e, C., et al. 2010, \aj, 139, 906
\bibitem[Wright et al.(2010)]{wri10} Wright, E. L., Eisenhardt, P. R. M., Mainzer, A. K., et al. 2010, \aj, 140, 1868
\bibitem[Wu et al.(2015)]{wu15} Wu, X.-B., Wang, F., Fan, X., et al. 2015, Nature, 518, 512
\bibitem[Yang et al.(2016)]{yan16} Yang, J., Wang, F., Wu, X.-B., et al. 2016, \apj, 829, 33
\bibitem[Yang et al.(2019)]{yan18} Yang, J., Wang, F., Fan, X., et al. 2019, \apj, 871, 199
\bibitem[Yip et al.(2004)]{yip04} Yip, C. W., Connolly, A. J., Vanden Berk, D. E., et al. 2004, \aj, 128, 2603
\bibitem[York et al.(2000)]{yor00} York, D. G., Adelman, J., Anderson, Jr., J. E., et al. 2000, \aj, 120, 1579
\end{thebibliography}
\end{document}